\documentclass{JHEP3}
\usepackage{amsmath}
\input epsf
\usepackage{epsfig}
\usepackage{amssymb}
\usepackage{graphics}
\usepackage[active]{srcltx}

\newcommand{\be}{\begin{equation}}
\newcommand{\ee}{\end{equation}}
\newcommand{\bea}{\begin{eqnarray}}
\newcommand{\eea}{\end{eqnarray}}
\newcommand{\bean}{\begin{eqnarray*}}
\newcommand{\eean}{\end{eqnarray*}}
\newcommand{\nn}{\nonumber \\}

\def\W #1{\widetilde{#1}}
\def\WH #1{\widehat{#1}}

\def\eref#1{(\ref{#1})}

\def\a{{\alpha}}

\def\eps{\epsilon}

\def\Label#1{\label{#1}%
  \smash{\hbox to0pt{\raise1ex\hbox{\tiny[#1]}\hss}}}

\preprint{2015}
\title{{\LARGE {CHY-construction of Planar Loop Integrands of  Cubic Scalar Theory }}}
\author{{\normalsize \mbox{Bo~Feng\mbox{$^{1,2}$}}}\\
\mbox{{\mbox{$^1$}\ Zhejiang Institute of Modern Physics}}\\
\mbox{{\ \;\! Physics Department, Zhejiang University}}\\
\mbox{{\ \;\! PR China, 310027}}
\medskip \\
\mbox{{\mbox{$^2$}\ Niels Bohr International Academy and Discovery Center}}\\
\mbox{{\ \;\! The Niels Bohr Institute, University of Copenhagen, }}\\
\mbox{{\ \;\! Blegdamsvej 17, DK-2100 Copenhagen, Denmark}}}
\date{\today}
\abstract{ In this paper, by treating massive loop momenta to massless momenta in higher dimension,
we are able to treat all-loop scattering equations as tree ones.
As an application of the new aspect, we consider the CHY-construction of bi-adjoint $\phi^3$ theory. We
present the explicit formula for two-loop planar integrands. We discuss carefully how to
subtract various forward singularities in the construction. We count the number of terms obtained
by our formula and by direct Feynman diagram calculation and find the perfect match, thus provide a strong support for our results.
}


\begin{document}

\section{Introduction}

The discovery of CHY-formula for tree-level scattering amplitudes by  Cachazo,
He and Yuan [CHY] in a series of papers
\cite{Cachazo:2013gna,Cachazo:2013hca, Cachazo:2013iea,
Cachazo:2014nsa,Cachazo:2014xea} has provided a novel way to calculate and
understand scattering amplitudes. In this construction, a set of algebraic equations ( called the {\bf scattering equations}) has played a crucial role. These equations appear in the literature in a variety of contexts
\cite{Fairlie:1972abc,Roberts:1972abc,Fairlie:2008dg,Gross:1987ar,Witten:2004cp,
Caputa:2011zk,Caputa:2012pi,Makeenko:2011dm,Cachazo:2012uq}. More explicitly,  scattering equations of $n$-particles are given by
\bea {\cal S}_a\equiv \sum_{b\neq a} {2 k_a\cdot k_b\over z_{ab}},~~~~z_{ab}\equiv z_a-z_b,~~~~a=1,2,...,n~,~~~\label{SE-def}\eea
where the $z_a$ is the location of $a$-th particle in the Riemann surface.
Although there are $n$ equations, only $(n-3)$ of them are independent, which can be seen by following three identities among them:
\bea \sum_a {\cal S}_a=0,~~~~\sum_a {\cal S}_a z_a =0,~~~~\sum_a {\cal S}_a z_a^2=0,~~~~\label{SE-rel}\eea
under the  momentum conservation and null conditions $k_a^2=0$. The tree-level amplitude
is calculated by following formula
\bea {\cal A}_n& = & \int {\left(\prod_{a=1}^n dz_a\right)\over d\omega}  \Omega({\cal S}) {\cal I},~~~~d\omega={d z_r d z_s d z_t\over z_{rs} z_{st} z_{tr}} ~~~~\label{gen-A-1}\eea
where ${\cal I}$ is the so called {\bf CHY-integrand} and $d\omega$ is the volume of $SL(2,C)$ group, where we have used the symmetry to fix locations of three variables $z_{r}, z_{s}, z_t$. The $\Omega({\cal S})$ is given by
\bea \Omega({\cal S})_{ijm} & = &
z_{ij}z_{jm}z_{mi} \prod_{a\neq i,j,k,m}\delta\left( {\cal S}_a\right)~~~~\label{SE-fix}\eea
where $(n-3)$-independent delta-functions of scattering equations have been imposed. Since there are $(n-3)$ variables and $(n-3)$ equations, there is no integration left to do in \eref{gen-A-1}. For each solution of delta-functions, we get a result after inserting it into the CHY-integrand ${\cal I}$. The amplitude is given by summing over all $(n-3)!$ results.

The correctness of CHY-formula has been understood from various points of view. In \cite{Dolan:2013isa}, using the BCFW on-shell recursion
relation \cite{Britto:2004ap,Britto:2005fq} the validity of the CHY
construction for $\phi^3$ theory and Yang-Mills theory has been
proved. Using ambitwistor string
theory
\cite{Mason:2013sva,Berkovits:2013xba,Gomez:2013wza,Adamo:2013tsa,Geyer:2014fka,
Geyer:2014lca,Casali:2014hfa,Adamo:2015hoa,Casali:2015vta,Ohmori:2015sha,Geyer:2015bja},
by calculating corresponding correlation functions of different world-sheet fields,
different  CHY-formulas for different theories have been derived alongside with the natural appearance
of  scattering equations. In \cite{Bjerrum-Bohr:2014qwa},  inspired by the field theory
limit of string theory, a dual model has been introduced. Using this idea,  a direct connection between the CHY-formula and the standard tree-level
Feynman diagrams  has been established in \cite{Baadsgaard:2015voa, Baadsgaard:2015ifa}.

The CHY-formula (or CHY-construction) has divided  calculating scattering amplitudes of a given theory into two
parts: (a) finding solutions of scattering equations and (b) finding the corresponding  CHY-integrand ${\cal I}$, which is the rational functions of locations $z_a$ for the given theory. Among these two parts, the former task is universal for all theories while the later task does  depend on the detail of theories.
Although there are some general principles to guide the construction of CHY-integrands, we still do not know the general algorithm for all theories. However, amazing progress has been made in \cite{Cachazo:2014xea} where integrands are known for many theories.

Although looks simple, scattering equations are not so easy to solve. By proper transformation, scattering equations become a set of algebraic equations as shown in \cite{Dolan:2014ega}. From this aspect, several work has appeared \cite{Kalousios:2015fya, Huang:2015yka,Sogaard:2015dba, Dolan:2015iln, Cardona:2015ouc, Cardona:2015eba} by exploring the powerful
computational algebraic geometry method, such as the companion matrix, the Bezoutian matrix, the elimination theorem. A different approach is given in \cite{Cachazo:2015nwa} by mapping the problem
to the known result of bi-adjoint $\phi^3$ theory. Using the generalized KLT relation and  Hamiltonian decompositions of certain 4-regular graphs, one can bypass solving  scattering equations and read out results directly. Another powerful method is given in \cite{Baadsgaard:2015voa, Baadsgaard:2015ifa},
where a mapping rule between CHY-integrands and  tree-level Feynman diagrams has been given. In this paper, we will use the mapping rule heavily and related results have  been given in the Appendix A.

Encouraged by the success at tree-level, a lot of efforts have been done to generalize to loop-level \cite{Adamo:2013tsa, Casali:2014hfa, Adamo:2015hoa}. A breakthrough is given in \cite{Geyer:2015bja} by Geyer, Mason, Monteiro and Tourkine~\cite{Geyer:2015bja}. They show
how to reduce the problem of genus one to a modified problem on the Riemann sphere, where the analysis is essentially as
at tree-level. Using the picture, they provide a conjecture
to any loop order. In \cite{Baadsgaard:2015hia, He:2015yua}, the one-loop integrand of bi-adjoint $\phi^3$ theory has been proposed, while in \cite{Geyer:2015jch,Cachazo:2015aol} more general theories such as Yang-Mills theory and gravity theory have been treated at the one-loop level. Among these results, the generalization of mapping rule to one-loop level given  in \cite{Baadsgaard:2015hia}
will be very useful. In fact, in this paper, we will show that this mapping rule could be generalized to all loops.

In this paper, we will generalize above one-loop results to higher loops. We will write down all loop scattering equations. The key idea of our approach is to treat  massive loop momenta as  massless momenta in a higher dimension. Using the idea, we effectively reduce the loop problem to
tree one. In fact, the same idea has been explored by the ${\cal Q}$-cut construction in \cite{Baadsgaard:2015twa,Huang:2015cwh}. After having loop scattering equations, we construct the CHY-integrand, which will produce two-loop planar integrand of bi-adjoint $\phi^3$ theory.

The plan of the paper is following. In the section two, we have reviewed the mapping rule between  CHY-integrands and Feynman diagrams of bi-adjoint $\phi^3$ theory and discussed how to write down CHY-integrands for tree diagrams with a given set of poles. In the section three, we discuss all loop scattering equations. In the section four, we construct the two-loop CHY-integrand for $\phi^3$ theory. To carry out the
construction, we have carefully discussed related forward singularities when sewing tree to become loops and how to remove them. In the section five, by the matching of the number of  terms obtained by CHY-construction and by Feynman diagrams, we provide a strong support for our result. In the section six, a brief conclusion is given.

\section{Tree-level amplitude of color ordered bi-adjoint $\phi^3$ theory}

In this part, we will review relevant results of color ordered bi-adjoint $\phi^3$ theory at tree-level, especially the
mapping rule between  tree-level Feynman diagrams and  tree-level CHY-integrands. Using this mapping rule, we can discuss how to remove certain Feynman diagrams from a given CHY-integrand. Before doing so, let us define following compact notation
\bea  [i_1,i_2,\ldots,i_m]\equiv \sum_{1\leq a<b \leq m} 2k_{i_a}\cdot k_{i_b}~.~~\label{notation-1} \eea

\begin{figure}
  \includegraphics[width=15cm]{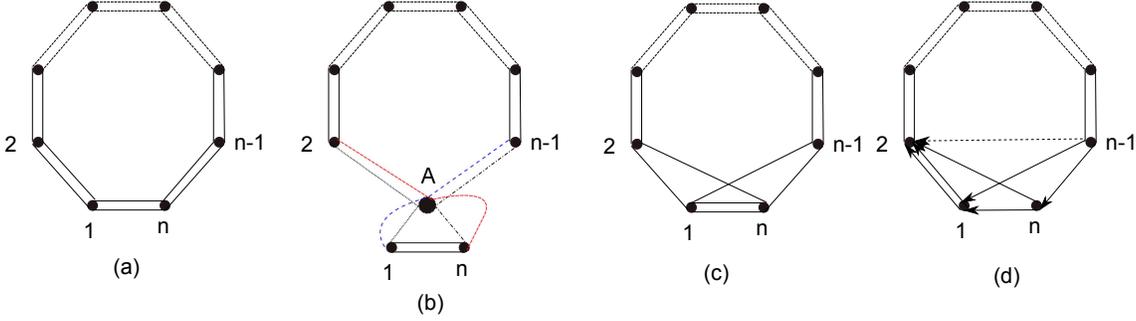}\\
  \caption{
The CHY-graph for Feynman diagrams with pole $s_{n1}$. (a) The CHY-graph for full  $n$-point tree-level amplitude; (b) The "pinching" picture where the new vertex $A$ represents the combination of vertexes $1,n$. (c) The  CHY-graph for Feynman diagrams containing pole $s_{n1}$ obtained from (b) after lifting the $(n-1)$-point graph to the $n$-point graph. (d) The CHY-graph after subtracting (c) from (a), where we have used  arrows to indicate the
direction.
  } \label{n-tree}
\end{figure}

Now we discuss the mapping rule given in \cite{Baadsgaard:2015voa, Baadsgaard:2015ifa}. First it is worth to notice that by Mobius invariance  each factor $z_i$ should have degree $-4$ in the CHY-integrand, thus one can represent the CHY-integrand by a graph, where each factor $z_{ij}\equiv (z_i-z_j)$ in the denominator corresponds to one (arrowed) solid line connecting vertexes $i,j$ and each factor $z_{ij}$ in the numerator  corresponds to one (arrowed) dash line connecting vertexes $i,j$.
Such graph will be called the CHY-graph. Given a CHY-integrand (or CHY-graph), the result obtained from CHY-formula will  a sum of inverse-products of multi-index Mandelstam invariants denoted $s_{i_1 ... i_m}=(k_{i_1}+...+k_{i_m})^2
=[i_1,i_2,\ldots,i_m]$ when all $k_i^2=0$, i.e.,
\bea \prod_{a=1}^{n-3}{1/s_{P_a}}=\prod_{a=1}^{n-3}{1\over [P_a]},\vspace{-5pt}\label{tree-pole}\eea
for $n$-point tree-level amplitudes. Each $P_a\!\subset\!\{1,\ldots,n\}$ denotes a subset of legs that we can always take to have at most $n/2$ elements (because $s_{P}\!\!=\!s_{P^\complement}$, with $P^\complement\!\equiv\!\mathbb{Z}_n\backslash P$, by momentum conservation). The collections of subsets $\{P_a\}$ appearing in (\ref{tree-pole}) must satisfy the following criteria:
\begin{itemize}
\item for each pair of indices $\{i,j\}\!\subset\!P_a$ in each subset $P_a$, there are exactly $(2|P_a|-2)$ factors of $(z_i-z_j)$ appearing in the denominator of $\mathcal{I}(z_1,\ldots,z_n)$;\\[-24pt]

\item each pair of subsets $\{P_a,P_b\}$ in the collection is either nested or complementary---that is, $P_a\!\subset\!P_b$ or $P_b\!\subset\!P_a$ or $P_a\!\subset\!P_b^\complement$ or $P_b^\complement\!\subset\!P_a$;\\[-24pt]
    \end{itemize}
If there are no collections of $(n-3)$ subsets $\{P_a\}$ satisfying the criteria above, the result of integration will be zero. One simple example using above rule is that
\bea \frac{1}{z_{12}^2 z_{23} z_{34}z_{45}z_{15}z_{35}^2z_{14}z_{24}} \Longleftrightarrow  {1\over s_{12} s_{35}}\eea
Another important example is the CHY-integrand for the full tree-level amplitude of $\phi^3$ theory with ordering $\{1,2,...,n\}$ (the corresponding CHY-graph is given by the diagram (a) in the Figure \ref{n-tree})
\bea {\cal I}^{CHY}_{n}(\{1,2,...,n\})= {1\over z_{12}^2 z_{23}^2 ...z_{(n-1)n}^2 z_{n1}^2}~.~~~\label{phi3-full}\eea
There is one fundamental formula, which will be useful later:  the number of color ordered $n$-point tree-level Feynman diagrams of  $\phi^3$ theory is given by
\bea C(n)= {2^{n-2}(2n-5)!!\over (n-1)!}~.~~~~\label{count-tree-n}\eea

Having presented the rule above, we try to find the CHY-integrand which gives Feynman diagrams of certain type, such as these in the Figure \ref{1loop-exl} and the Figure \ref{Fe-cut}.
Let us start with the simplest case, i.e., the (B-2) type of Figure \ref{1loop-exl}, where  we assume that  $1,n$ are always attached to the same cubic vertex and then they combine together to connect to other legs. If we cut the propagator $s_{1n}$, we will be left with color ordered full tree-level amplitude with $(n-1)$-legs. This picture motivates us an operation called the "{\bf pinching}" where vertexes $1,n$ are combined to become a new vertex $A$ (see the diagram (b) in the Figure \ref{n-tree}). It is worth
to notice that in (b) we have drawn four lines in different colors and styles to emphasize when we lift the $(n-1)$-point graph to $n$-point graph, how these lines are connected. Also the group $A,1,2$ itself is the CHY-graph corresponding to the expression \eref{phi3-full} with $n=3$. The lift graph of (b) is given in the diagram (c) in the Figure \ref{n-tree}. When translating above manipulation at the graph level to expression, we find that
\bea {\cal I}_{n;s_{n1}}^{CHY}(\{1,2,...,n\})= {1\over  z_{23}^2 ...z_{(n-2)(n-1)}^2\left(z_{(n-1)n}z_{(n-1)1}\left( z_{n1}^2\right) z_{n2} z_{12}\right)}~~~~\label{phi3-1n-pole}\eea
Using the mapping rule, one can check that the CHY-integrand \eref{phi3-1n-pole} will give expression contains $C(n-1)$ terms with the fixed pole $s_{n1}$ (see Eq.\eref{count-tree-n}), which is the right counting number. Now it is obvious that if we want
to remove these Feynman diagrams of the (B-2) type, we should  subtract the CHY-integrand \eref{phi3-1n-pole} from the CHY-integrand \eref{phi3-full} and  get
\bea & &{1\over z_{23}^2 ... z_{(n-2)(n-1)}^2}\left\{ {1\over z_{(n-1)n}^2 z_{n1}^2 z_{12}^2}-{1\over z_{(n-1)n}z_{(n-1)1} z_{n1}^2 z_{n2} z_{12}}\right\}\nn &= &  {1\over z_{23}^2 ... z_{(n-2)(n-1)}^2}
{z_{(n-1)2}\over z^2_{(n-1)n}z_{(n-1)1} z_{n1} z_{n2} z^2_{12}}~~~\label{remove-s1n}\eea
where the explicit pole $z_{n1}^2$ in the denominator has been canceled. The final CHY-integrand can be represented by the
diagram (d) in the Figure \ref{n-tree}.

\begin{figure}
  \includegraphics[width=15cm]{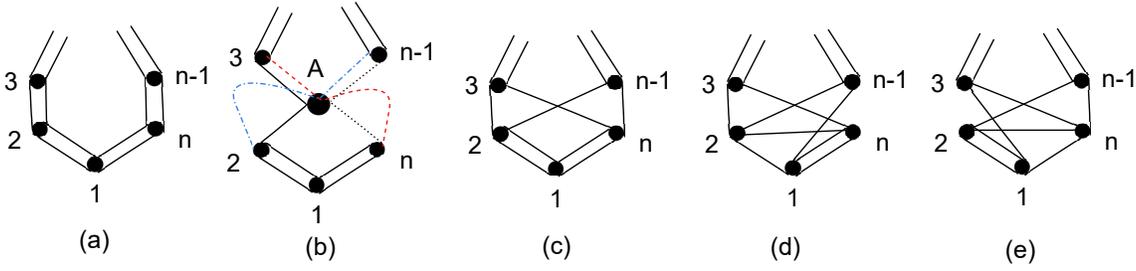}\\
  \caption{
 (a) The CHY-graph for full  $n$-point tree-level amplitude; (b) The "pinching" picture where the new vertex $A$ represents the combination of vertexes $n,1,2$. (c) The  CHY-graph for all Feynman diagrams with pole $s_{n12}$ obtained from (b) after lifting the $(n-2)$-point graph to the $n$-point graph. (d) The CHY-graph having the fixed poles $s_{n12}$ and $s_{n1}$; (e) The CHY-graph having the fixed poles $s_{n12}$ and $s_{12}$;
  } \label{polen12}
\end{figure}

Having done the simplest case, now we consider the CHY-integrand, which produces all Feynman diagrams containing a given pole, for example, $s_{n12}$. Again we can pinch three vertexes $n,1,2$ together to reduce $n$-legs to $(n-2)$-legs (see the diagram (b) in Figure \ref{polen12}). After lifting we get the
CHY-graph (see the diagram (c) in Figure \ref{polen12}), which contains all Feynman diagrams having the pole $s_{n12}$. The corresponding CHY-integrand is obtained by replacing
$z_{(n-1)n}^2 \left(z_{n1}^2 z_{12}^2\right) z_{23}^2$ in the denominator of \eref{phi3-full} to the factor $z_{(n-1)n}z_{(n-1)2} \left(z_{n1}^2 z_{12}^2\right)z_{n3} z_{23}$, i.e.,
\bea {\cal I}^{CHY}_{n; s_{n12}}(\{1,2,...,n\})= {1\over  z_{34}^2 ...z_{(n-2)(n-1)^2}\left(z_{(n-1)n}z_{(n-1)2} \left(z_{n1}^2 z_{12}^2\right)z_{n3} z_{23}\right)}~~~~\label{phi3-n12-pole}\eea
If we subtract the CHY-integrand \eref{phi3-n12-pole} from the CHY-integrand \eref{phi3-full}, we will get
\bea & &{1\over z_{34}^2 ... z_{(n-2)(n-1)}^2}\left\{ {1\over z_{(n-1)n}^2 \left(z_{n1}^2 z_{12}^2\right) z_{23}^2}-{1\over z_{(n-1)n}z_{(n-1)2} \left(z_{n1}^2 z_{12}^2\right)z_{n3} z_{23}}\right\}\nn &= &  {1\over z_{34}^2 ... z_{(n-2)(n-1)}^2}
{z_{(n-1)3}z_{n2}\over z^2_{(n-1)n}z_{(n-1)2} \left(z_{n1}^2 z_{12}^2\right)z_{n3} z^2_{23}}~.~~\label{remove-sn12}\eea
Using the "pinching" operation above pattern can be easily generalized to find the CHY-integrand which produces all Feynman diagrams containing a given pole, for example, $s_{n12..k}$. What we need to do is following replacement of
the factor $z_{(n-1)n}^2 \left(z_{n1}^2 z_{12}^2... z_{(k-1)k}^2\right) z_{k(k+1)}^2$ in the denominator of \eref{phi3-full} to the factor $z_{(n-1)n} z_{(n-1)k} \left(z_{n1}^2 z_{12}^2... z_{(k-1)k}^2\right)z_{n(k+1)} z_{k(k+1)}$, i.e.,
\bea {1\over ...z_{(n-1)n}^2 \left(z_{n1}^2 z_{12}^2... z_{(k-1)k}^2\right) z_{k(k+1)}^2...}
\Longrightarrow {1\over ... z_{(n-1)n} z_{(n-1)k} \left(z_{n1}^2 z_{12}^2... z_{(k-1)k}^2\right)z_{n(k+1)} z_{k(k+1)}...}~.~\label{remove-single-pole} \eea
Above  replacement rule can be nicely represented as following: {\sl for each fixed pole $s_{n12..k}$ we multiply  by a corresponding  factor
\bea {\cal P}[n-1,n,k,k+1]\equiv { z_{(n-1) n} z_{k(k+1)} \over z_{(n-1)k} z_{n(k+1)}}~~~~~\label{single-pole-rep}\eea
where $n,k$ as the first and the last legs in the ordering of the  specified  pole}.

Having observed the pattern, now it is easy to write down the corresponding CHY-integrand with a given pole structure (we will call it as  the "signature"). Let us give a few examples:
\begin{itemize}

\item With fixed poles $s_{n12}$ and
$s_{456}$, the integrand is given by ${\cal I}^{CHY}_{n}(\{1,2,...,n\}){\cal P}[n-1,n,2,3] {\cal P}[3,4,6,7]$.

\item With fixed poles $s_{12}$ and  $s_{34}$, the integrand is given by ${\cal I}^{CHY}_{n}(\{1,2,...,n\}){\cal P}[n,1,2,3] {\cal P}[2,3,4,5]$. It is worth to notice that
     $z_{23}^2$ in numerator will cancel the  $z_{23}^2$ in denominator of ${\cal I}^{CHY}_{n}$.

\item With fixed poles $s_{n1}$ and  $s_{n12}$, the integrand is given by ${\cal I}^{CHY}_{n}(\{1,2,...,n\}){\cal P}[n-1,n,1,2] {\cal P}[n-1,n,2,3]$. For this case, pole $s_{n1}$ is inside the pole $s_{n12}$.

\end{itemize}

Above examples have only two poles and it is easy to check that the numerator of the final expression is one. Thus we can check our claim easily using the mapping rule \eref{tree-pole}. However, when we fix three or more poles, some interesting thing happens: the numerator of final expression could be nontrivial. For example, with the signature $s_{12} s_{123} s_{1234}$ at eight points, after applying our rule \eref{single-pole-rep} we get
\bea
{z_{81}\over z_{12}^2 z_{23} z_{34} z_{45} z_{56}^2 z_{67}^2 z_{78}^2  z_{84} z_{15} z_{83} z_{14}z_{82} z_{13}}~.~~\label{Bound-3step}\eea
Since the numerator is not one, we could not apply the mapping rule directly. To solve the problem, we need to use following identity
\bea {z_{ab} z_{dc}\over z_{ac} z_{bc}} ={ z_{ad}\over z_{ac}}-{z_{bd}\over z_{bc}}~.~~~\label{z-identity}\eea
Applying to our case with $z_{ab}=z_{81}$, we need to find $c,d$. There are some conditions. First the degree of factor $z_{dc}$ in the denominator of  original expression can only be zero or one. Thus, after multiplying $1={z_{dc}\over z_{dc}}$ and then applying identity \eref{z-identity}, we will not end up with factor $z_{dc}$ having degree more than two in denominator. Secondly, we should require
the original expression has factors $z_{ac}, z_{bc}$ in denominator to give the left handed side of
\eref{z-identity}. Finally, we should require
the original expression has factors $z_{ad}$ and $z_{bd}$ in denominator to cancel the corresponding factor in numerator appearing after using \eref{z-identity}. If we can find such $d,c$, we can reduce
the problem to trivial one and then apply our mapping rule. For the example given in \eref{Bound-3step}, it is easy to see that $d,c$ can be chosen from $\{2,3,4\}$. In fact, there are six possible choices and we have checked each one.  With the choice $c=2,d=3$ we have
\bea & &{z_{81}\over z_{12}^2 z_{23} z_{34} z_{45} z_{56}^2 z_{67}^2 z_{78}^2  z_{84} z_{15} z_{83} z_{14}z_{82} z_{13}}={-1\over z_{12} z_{23}^2 z_{34} z_{45} z_{56}^2 z_{67}^2 z_{78}^2  z_{84} z_{15} z_{83} z_{14} z_{13}} \times { z_{81} z_{32} \over  z_{82} z_{12}}\nn
& = & {-1\over z_{12} z_{23}^2 z_{34} z_{45} z_{56}^2 z_{67}^2 z_{78}^2  z_{84} z_{15} z_{83} z_{14} z_{13}}
\left( {z_{83}\over z_{82}}-{z_{13}\over z_{12}}\right)\nn
& = &{-1\over z_{12} z_{23}^2 z_{34} z_{45} z_{56}^2 z_{67}^2 z_{78}^2  z_{84} z_{15} z_{82} z_{14} z_{13}}+{+1\over z_{12}^2 z_{23}^2 z_{34} z_{45} z_{56}^2 z_{67}^2 z_{78}^2  z_{84} z_{15} z_{83} z_{14} } ~.~~\label{3pole-decom} \eea
Using the mapping rule, we can calculate each term and sum them up. It is easy to check that they
indeed give all terms having above signature.

We can continue to more complicated examples, for example, the one with signature $s_{12} s_{123} s_{1234} s_{12345}$ at eight points. One more pole means to  multiply another factor
\bea & & {z_{81}\over z_{12}^2 z_{23} z_{34} z_{45} z_{56}^2 z_{67}^2 z_{78}^2  z_{84} z_{15} z_{83} z_{14}z_{82} z_{13}}\times { z_{81} z_{56}\over z_{85} z_{16}}\nn
& = & \left( {-1\over z_{12} z_{23}^2 z_{34} z_{45} z_{56}^2 z_{67}^2 z_{78}^2  z_{84} z_{15} z_{82} z_{14} z_{13}}+{+1\over z_{12}^2 z_{23}^2 z_{34} z_{45} z_{56}^2 z_{67}^2 z_{78}^2  z_{84} z_{15} z_{83} z_{14} }\right)\times { z_{81} z_{56}\over z_{85} z_{16}}~~~\label{Bound-4step}\eea
where we have used the result \eref{3pole-decom}. Now we use similar idea to do decomposition of these two terms.
For the first term we take $c=2,d=4$ and obtain
\bea
&  & {- 1 \over  z_{23}^2 z_{34} z_{45} z_{56} z_{67}^2 z_{78}^2  z_{82} z_{15}  z_{14} z_{13} z_{85} z_{16}z_{42}}+ {+ 1 \over  z_{23}^2 z_{34} z_{45} z_{56} z_{67}^2 z_{78}^2  z_{84} z_{15}  z_{12} z_{13} z_{85} z_{16}z_{42}}~. \eea
For the second term, we take $c=5, d=4$  and  get
\bea {1\over z_{12}^2 z_{23}^2 z_{34} z_{45}^2 z_{56} z_{67}^2 z_{78}^2  z_{85} z_{83} z_{14} z_{16} }+ {-1\over z_{12}^2 z_{23}^2 z_{34} z_{45}^2 z_{56} z_{67}^2 z_{78}^2  z_{84} z_{83} z_{15} z_{16} }~.\eea
Using the mapping rule to above four terms and summing them up, we do get all terms having the signature of four fixed poles.

\section{All Loop scattering equations}

In this section, we will discuss  general $m$-th loop scattering equations. First we will review
 the construction given in \cite{Geyer:2015bja}, then we give another understanding of
 these equations from the point view of  higher dimension.
 To establish the relation between  $m$-th loop $n$-point scattering equations  and tree scattering equations of $(n+2m)$-points, we  use following convention:
$k_i$, $i=1,...,n$ for momenta of $n$ external legs, while $k_{n+2j-1}=-k_{n+2j}$ with $j=1,...,m$ for the $j$-th loop momentum. While we still impose $k_i^2=0$ for $i=1,...,n$,   loop momenta $k_{n+2j-1}$ are general massive.

To derive loop scattering equations, we start from the $m$-th loop one-form
\bea P=\sum_{r=1}^m  k_{n+2r-1} { (z_{n+2r-1}-z_{n+2r})dz\over (z-z_{n+2r-1})
(z-z_{n+2r})}+\sum_{i=1}^n k_i {dz\over z-z_i}~~~\label{mloop-P}\eea
where $z_i, i=1,..,n$ are marked points for external legs while $z_{n+2r-1}, z_{n+2r}$, $r=1,...,m$
are new marked points for pinching Riemann sphere.
It is easy to see that $P^2$ contains double poles, thus we define
\bea S(z) & =& P^2-\sum_{r=1}^m  k^2_{n+2r-1} { (z_{n+2r-1}-z_{n+2r})^2d^2z\over (z-z_{n+2r-1})^2
(z-z_{n+2r})^2}~, \eea
which contains only single poles at all marked points $z_i$,  $i=1,2...,n+2m$ . Calculating these residues, we get
    \bea {\cal S}_a & = & \sum_{j\neq a,1}^n {2k_a\cdot k_j\over z_a-z_j}+\sum_{t=1}^m
    \left({2 k_a\cdot k_{n+2t-1}\over z_a-z_{n+2t-1}}+{2 k_a\cdot k_{n+2t}\over z_a-z_{n+2t}}\right) ,~~~~1\leq a\leq n~~~\label{2nd-Ea-1}\eea
for $n$ external marked points  and
\bea {\cal S}_{n+2t-1} & = & \sum_{a=1}^n {  2k_{n+2t-1}\cdot k_a\over z_{n+2t-1}-z_a}
+\sum_{s=1,s\neq t}^m
    \left({2 k_{n+2t-1}\cdot k_{n+2s-1}\over z_{n+2t-1}-z_{n+2s-1}}+{  2k_{n+2t-1}\cdot k_{n+2s}\over z_{n+2t-1}-z_{n+2s}}\right)~,\nn
{\cal S}_{n+2t} & = & \sum_{a=1}^n {2  k_{n+2t}\cdot k_a\over z_{n+2t}-z_a}
+\sum_{s=1,s\neq t}^m
    \left({ 2k_{n+2t}\cdot k_{n+2s-1}\over z_{n+2t}-z_{n+2s-1}}+{2  k_{n+2t}\cdot k_{n+2s}\over z_{n+2t}-z_{n+2s}}\right),~~~~1\leq t\leq m~~~\label{2nd-Ea-2}    \eea
for new marked points corresponding to the $t$-th loop momentum. These $(n+2m)$ equations given in \eref{2nd-Ea-1} and \eref{2nd-Ea-2} are the $m$-th loop scattering equations we are looking for.

Now we compare these equations with the corresponding tree-level scattering equations of $(n+2m)$-points given by
\be {\cal S}_a\equiv \sum_{b\neq a} {2k_a\cdot k_b\over z_a-z_b},~~~~~a=1,2,...,n+2m.~~~~\label{SE-def}\ee
They are exactly same for $a=1,...,n$, except  the  remaining $2m$ momenta satisfying $k_{n+2j-1}=-k_{n+2j}$ (i.e.,
in the forward limit). However, for $a=n+1,...,n+2m$,
 terms like ${2k_{n+2t-1}\cdot k_{n+2t}\over
z_{n+2t-1}-z_{n+2t}}$ in tree-level scattering equations have been dropped in the $m$-th loop scattering equations. The dropping of these terms can, in fact, be traced back  to  the numerator $(z_{n+2r-1}-z_{n+2r})$ of the first term in \eref{mloop-P}. This difference is crucial as we will explain later.

Having obtained  loop scattering equations, let us check  their Mobius covariance. Under the  Mobius transformation $ z'={az+b\over cz +d}$, one find
\bea z_{ij}'=
{(ad-bc)\over (cz_i+d)(c z_j+d)} z_{ij}~,~~~\label{SLC-sigma}\eea
thus it is easy to check that  for ${\cal S}_{1\leq a\leq n}$ we have
\bea {\cal S}_a & \to  &{(c z_a+d)\over (ad-bc)} \left\{\sum_{j\neq a,1}^n {2k_a\cdot k_j(c z_j+d)\over z_a-z_j}+\sum_{t=1}^m
\left({2 k_a\cdot k_{n+2t-1}(c z_{n+2t-1}+d)\over z_a-z_{n+2t-1}}+{ 2k_a\cdot k_{n+2t}(c z_{n+2t}+d)\over z_a-z_{n+2t}}\right) \right\}\nn
& = & {(c z_a+d)\over (ad-bc)} \left\{\sum_{j\neq a,1}^n \left({2 k_a\cdot k_j(c z_a+d)\over z_a-z_j}- 2 k_a\cdot k_j\right) \right. \nn & & \left.  +\sum_{t=1}^m
\left({ 2k_a\cdot k_{n+2t-1}(c z_{a}+d)\over z_a-z_{n+2t-1}}-2 k_a\cdot k_{n+2t-1}+{2 k_a\cdot k_{n+2t}(c z_{a}+d)\over z_a-z_{n+2t}}-2k_a\cdot k_{n+2t}\right) \right\}\nn
& = & {(c z_a+d)^2\over (ad-bc)}{\cal S}_a~, \eea
and for ${\cal S}_{n< a\leq n+2m}$ we have
\bea
 {\cal S}_{n+2t-1} & \to  & {(c z_{n+2t-1}+d)\over (ad-bc)} \left\{\sum_{a=1}^n { 2 k_{n+2t-1}\cdot k_a(cz_a+d)\over z_{n+2t-1}-z_a}\right. \nn & & \left.
+\sum_{s=1,s\neq t}^m
    \left({2 k_{n+2t-1}\cdot k_{n+2s-1}(c z_{n+2s-1}+d)\over z_{n+2t-1}-z_{n+2s-1}}+{ 2 k_{n+2t-1}\cdot k_{n+2s}(c z_{n+2s}+d)\over z_{n+2t-1}-z_{n+2s}}\right)\right\}\nn
& = & {(c z_{n+2t-1}+d)\over (ad-bc)} \left\{\sum_{a=1}^n \left({ 2 k_{n+2t-1}\cdot k_a(cz_{n+2t-1}+d)\over z_{n+2t-1}-z_a}-2k_{n+2t-1}\cdot k_a\right)\right. \nn & & \left.
+\sum_{s=1,s\neq t}^m
    \left({ 2k_{n+2t-1}\cdot k_{n+2s-1}(c z_{n+2t-1}+d)\over z_{n+2t-1}-z_{n+2s-1}}-2k_{n+2t-1}\cdot k_{n+2s-1}\right.\right. \nn & & \left.\left.+{ 2 k_{n+2t-1}\cdot k_{n+2s}(c z_{n+2t-1}+d)\over z_{n+2t-1}-z_{n+2s}}-2k_{n+2t-1}\cdot k_{n+2s}\right)\right\}  \nn
& = &   {(c z_{n+2t-1}+d)^2\over (ad-bc)}  {\cal S}_{n+2t-1}    \eea
with similar expressions for $ {\cal S}_{n+2t}$.

The covariance indicates that there are three relations among these $(n+2m)$ scattering equations:
\bea \sum_{i=1}^{n+2m} {\cal S}_i=0,~~~~~\sum_{i=1}^{n+2m} z_i{\cal S}_i=0,~~~\sum_{i=1}^{n+2m} z_i^2 {\cal S}_i=0~.~~~\label{3-rel}\eea
We want to emphasize that in above calculations, we have used
only following three conditions: (1) massless condition $k_i^2=0$ for $i=1,...,n$; (2) momentum conservation $\sum_{i=1}^n k_i=0$; (3) forward limit $k_{n+2j-1}=-k_{n+2j}$ for $j=1,...,m$. In other words, we do not need to
impose $k_{n+2j-1}^2=0$, which is  one consequence of  dropped terms like ${2k_{n+2t-1}\cdot k_{n+2t}\over
z_{n+2t-1}-z_{n+2t}}$. In fact, it can be easily checked that without dropping these terms, the second and third relation in \eref{3-rel} can not be satisfied with above three conditions.

Now let us try to  understand the meaning  of dropping  terms like ${2k_{n+2t-1}\cdot k_{n+2t}\over
z_{n+2t-1}-z_{n+2t}}$. It is obviously that if $k_{n+2t-1}\cdot k_{n+2t}=-k_{n+2t}^2=0$, it will disappear
automatically. However, since they are loop momenta  we should not expect these conditions.
To make these two aspects consistent to each other, one nice idea is to use the {\it dimension reduction}. Let us  assume that all external momenta are in $D$-dimensional spacetime, then we can treat  massive
momenta in the $D$-dimensional spacetime to be  massless momenta in $(D+d)$-dimensional spacetime. This can be arranged because
 scattering equations are dimensional blind. In fact, using the idea, several groups have noticed that  scattering equations for massive particles\footnote{Other related works for massive particles can be found
in \cite{Dolan:2013isa,Naculich:2015zha,Naculich:2015coa,Lam:2015mgu}} at tree-level first proposed by Naculich in \cite{Naculich:2014naa} can be understood from this point of view. More explicit, let us rewrite the $(D+d)$-dimensional scattering equations as
\bea \sum_{j\neq i}{ \W k_i\cdot \W k_j\over z_i-z_j}=\sum_{j\neq i}{ k_i\cdot k_j+\Delta_{ij}\over z_i-z_j} \eea
where each $(D+d)$-dimensional momentum $\W k=k+\eta$  has been split into momentum $k$ in $D$-dimension and momentum $\eta$ in $d$-dimension, so $\Delta_{ij}=-\eta_i\cdot \eta_j$. It is easy to see that
$\Delta_{ij}=\Delta_{ji}$ and $\sum_{j\neq i}\Delta_{ij}=\eta_i\cdot \eta_i$ by momentum conservation in $d$-dimension. Thus massless condition $k_i^2-\eta_i^2=0$ in $(D+d)$-dimension gives the  mass $\sum_{j\neq i}\Delta_{ij}=m_i^2$
in $D$-dimension.

Above discussions lead us to a  new understanding of these $m$-th loop scattering equations in $D$-dimension:
{\it they are the tree-level scattering equations of $(n+2m)$-points under the forward limit, where $2m$'s momenta are massless in $(D+d)$-dimension while $n$ external momenta are massless in $D$-dimension}. An immediate
implication is that all contractions of the type $2 k_{n+2t-1}\cdot k_{n+2s-1}$ in \eref{2nd-Ea-2} should be understood as the contractions in $(D+d)$-dimension.

The new understanding of loop momenta in higher dimension has led an important application: since from the point of view of higher dimension they are massless, we have effectively cut $m$'s internal lines, so  $m$-th loop Feynman  diagrams open up to become  connected tree-level Feynman  diagrams. This idea has been used in \cite{Baadsgaard:2015hia} to construct  one-loop CHY integrands of $\phi^3$ theory (see also  \cite{He:2015yua,Geyer:2015jch, Cachazo:2015aol}). A more intensive application of reducing loop problems to tree-level ones has been demonstrated in the ${\cal Q}$-cut construction \cite{Baadsgaard:2015twa} (see also \cite{Huang:2015cwh}). In this paper, we will use the same idea to  write down  CHY loop integrands from corresponding tree ones.

Having understood the similarity and the connection with  tree-level cases, it is natural to
write down the integration formula for loop amplitudes as proposed in \cite{Geyer:2015bja}
\bea {\cal A}^D_{m-loop}=\int \prod_{i=1}^m {d^D \ell_i\over \ell_i^2} {\cal I}^{(D+d)}_{m-loop}~~~~\label{mloop}\eea
with
\bea {\cal I}^{(D+d)}_{m-loop}=\int {\left(\prod_{i=1}^{n+2m} dz_i\right)\over d\omega}\left(z_{ij}z_{jk}z_{ki} \prod_{a\neq i,j,k}^{n+2m}\delta\left( {\cal S}_a\right)\right) {\cal I}^{CHY}~.~~~\label{mloop-integral}\eea
Let us give some explanations for \eref{mloop} and \eref{mloop-integral}. First although loop momenta in \eref{mloop} are in $D$-dimension, when we use the CHY-formula to calculate ${\cal I}^{(D+d)}_{m-loop}$ as given in \eref{mloop-integral}, we should treat loop momenta as massless in $(D+d)$-dimension as explained above.   Thus we use  notation ${(D+d)}$ to emphasize this point. Secondly, the formula \eref{mloop-integral} is the familiar tree-level CHY
formula with $(n+2m)$-points, where
$d\omega={dz_r dz_s dz_t\over z_{rs}z_{st}z_{tr}}$ comes from gauge fixing of three locations of $z$'s by $SL((2,C)$ symmetry. While other part is universal for all theories, the CHY-integrand ${\cal I}^{CHY}$ is the one distinguishing different theories. Thus our main focus will be the construction of ${\cal I}^{CHY}$.

The construction of CHY-integrands needs to satisfy some constraints. One of the most important constraints is the Mobius invariance. To compensate the variation of measure part in \eref{mloop-integral}, under  the $SL(2,C)$ transformation, ${\cal I}^{CHY}$  should have  following transformation property
\bea {\cal I}^{CHY}\to \left(\prod_{i=1}^{n+2m} {(ad-bc)^2\over (c z_i+d)^4} \right)^{-1}{\cal I}^{CHY}~.~~\label{I-CHY-cond-1}\eea
A nice way to satisfy above transformation property is to construct various combinations carrying different weights
as demonstrated in \cite{Cachazo:2013gna,Cachazo:2013hca, Cachazo:2013iea, Cachazo:2014nsa,Cachazo:2014xea}. Two familiar factors with weight two are (more factors can be found in \cite{Cachazo:2014nsa,Cachazo:2014xea})
\bea C_{U(N)}(z)=\left( \sum_{\a\in S_n/Z_n} {{\rm Tr}(T^{\a(1)} ... T^{\a(n)})\over
z_{\a(1)\a(2)} ...z_{\a(n-1)\a(n)} z_{\a(n)\a(1)}}\right),~~~~E(\eps,k,z)=({\rm Pf}' \Psi(k,\eps,z) .~~~~\label{CHY-CE}\eea
Besides the weight conditions, there are other physical considerations, such as the soft limit, the factorization property etc.

\begin{figure}
  \includegraphics[width=15cm]{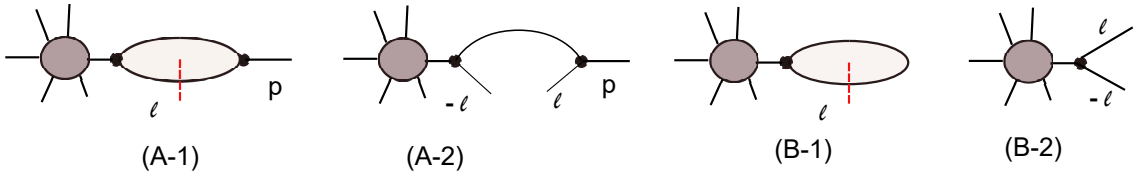}\\
  \caption{The excluded one-loop Feynman diagrams of $\phi^3$ theory and their corresponding trees after the cut}\label{1loop-exl}
\end{figure}

Although using the idea of dimension reduction, we have mapped the loop problem to tree one in \eref{mloop-integral}, the CHY-integrand ${\cal I}^{CHY}$ are not
exactly the tree-level CHY integrands we are familiar with. There are two main reasons. The first one is that
since  tree-level Feynman diagrams are obtained by cutting internal lines, there are many choices of which lines have been cut, thus one needs to sum over all allowed insertions of $2m$ extra legs (and possible  summing over polarization states of extra legs  if  particles running along the loop are not scalars). This phenomenon has been
discussed for one-loop cases in \cite{Baadsgaard:2015hia,He:2015yua,Geyer:2015jch, Cachazo:2015aol}. The second reason is more crucial: {\sl after cutting  loop diagrams to trees, we do not get all allowed tree-level diagrams of $(n+2m)$-points}. For example, for one-loop amplitude of massless theories, there are two kinds of diagrams we need to exclude: the tadpole diagrams (B-1) in Figure \ref{1loop-exl} and the massless bubble diagrams (A-1) in Figure \ref{1loop-exl}. After reducing loop diagrams to trees, we should exclude these diagrams (A-2), (B-2) in Figure \ref{1loop-exl} from allowed tree-level diagrams. These two kinds of tree diagrams are singular under the forward limit. Thus the true CHY-integrand in \eref{mloop-integral} should be the one from trees after subtracting these divergent parts.

However, the subtracting of these singular parts is very nontrivial. For some theories, for example, the supersymmetric theory, it has been shown in \cite{CaronHuot:2010zt} that
the singular forward limit disappears by supersymmetry\footnote{For other massless theories,
recent ${\cal Q}$-cut construction in \cite{Baadsgaard:2015twa} has given a way to remove forward singularities by using the scale deformation.}, so we do not need to worry about it. However, for pure Yang-Mills theory, the subtracting in the CHY frame is not completely clear. Since these subtleties, in this paper  we will focus on  planar loop integrands of color ordered bi-adjoint scalar $\phi^3$ theory. Although the theory is simple, it is good enough for our
one main purpose of the paper, i.e., to find the generalization of powerful mapping rule between CHY-integrands
and  Feynman diagrams given in \cite{Baadsgaard:2015voa, Baadsgaard:2015ifa,Baadsgaard:2015hia} at the tree-level and one-loop level.

\section{Two-loop CHY-integrand of $\phi^3$ theory}

Having discussed  all loop scattering equations, now we discuss how to write down all loop CHY-integrands in  \eref{mloop-integral}, at least for planar part of color ordered bi-adjoint $\phi^3$ theory. For simplicity, we will use the two-loop example to demonstrate our strategy, but the idea should be easily generalized to all loops. The key strategy to  loop CHY-integrands is to use the mapping rule found in papers \cite{Baadsgaard:2015voa, Baadsgaard:2015ifa,Baadsgaard:2015hia}. Using the mapping rule, if we know what is expressions from field theory side through Feynman diagrams, we could find the corresponding CHY-integrands.

\subsection{Analysis of two-loop  Feynman diagrams}

\begin{figure}
  \includegraphics[width=15cm]{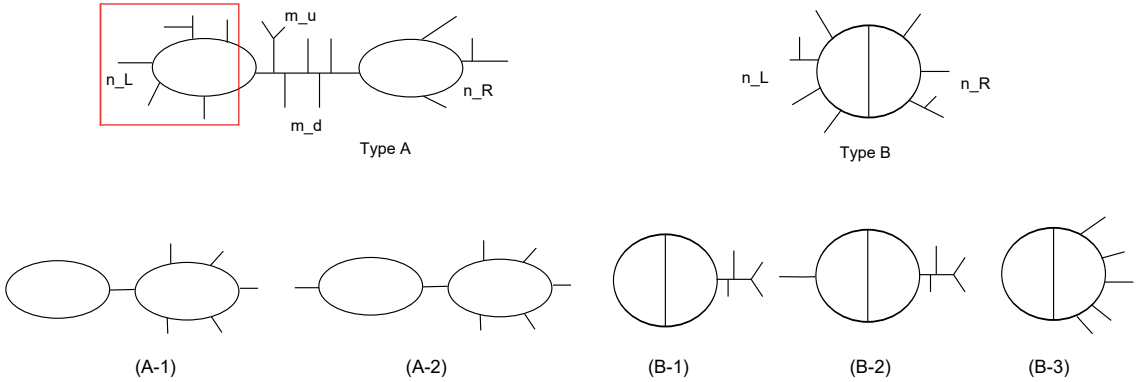}\\
  \caption{General planar two-loop Feynman diagrams Type (A) and Type (B) of $\phi^3$ theory. There  are some special two-loop diagrams: (A-1) one-loop tadpole; (A-2) one-loop massless bubble; (B-1) two-loop tadpole; (B-2) two-loop massless bubble; and (B-3) Reducible two-loop diagrams.}\label{Fe-1}
\end{figure}

Having above strategy, now we start to analyze  color ordered two-loop planar integrands obtained from Feynman diagrams. To have a clear picture of these integrands, let us start with the classification of  planar two-loop Feynman diagrams of $\phi^3$ theory. It is easy to see that  all diagrams are  divided into two types, i.e., type (A) and type (B) (see Figure \ref{Fe-1} ). The  type (A) is the relative trivial one as it is  given by two sub-oneloop diagrams. For these diagrams, we will use
$T_{(n_L; m_u,m_d; n_R)}$ to denote them, where $n_L, n_R, m_u, m_d$ are number of external legs attached to the left sub-oneloop, right sub-oneloop, the upper part of the middle line and the lower part of middle line. The type (B) is the really nontrivial two-loop diagram  with one mixed propagator. For these diagrams, we will use   $T_{(n_L; n_R)}$ to denote them, where $n_L, n_R$ are numbers of external legs attached to the left part and  right part.

Among these diagrams given in Figure \ref{Fe-1}, there are some singular two-loop diagrams, for which we will have more careful discussions. They are (see Figure \ref{Fe-1}):
\begin{itemize}

\item   When one of $n_L$ or $n_R$ of Type (A) is zero, we have one-loop tadpole structure as given by (A-1).

\item  When one $n_L$ or $n_R$ of Type (A) is one and all other external legs are grouped together to attach to
    another loop by only one vertex, we have one-loop massless bubble structure as given by (A-2).

\item  When $n_L$ or $n_R$ of Type (B)  are zero , we  get the reducible two-loop structure as given by (B-3). For the case (B-3), when all external legs are grouped together to attach to
    the loop by only one vertex, we get the two-loop tadpole structure as given by (B-1).

\item  When one of $n_L$ and $n_R$ of Type (B) is one and all other external legs are grouped together to attach to
    another loop by only one vertex, we get the two-loop massless bubble structure as given by (B-2).

\end{itemize}

From general two-loop Feynman diagrams, we see that two-loop integrands should be the sum of terms of  following two types\footnote{Under our convention,  the color ordering is clockwise. All external momenta are incoming while   loop momenta will run  along  clockwise direction, so when we cut inner propagator $\ell$ between the leg $1$ and the leg $2$, we should have the ordering  as $(1,-\ell, \ell,2)$, i.e, {\it moving along the clockwise direction is translated to moving from the left to the right}. Furthermore, the nontrivial mixed propagator will have the momentum $(\ell_1-\ell_2+R)^2$.  } (see the diagrams (A) and (B) in Figure \ref{Fe-1})
\bea {\cal I}_{A} & = & {1\over {\cal E} \left(\prod_{i} (\ell_1+ K_{i})^2\right)\left(\prod_{s} (\ell_2+ K_{s})^2\right)}\nn
{\cal I}_{B} & = & {1\over  {\cal E} \left(\prod_{i} (\ell_1+ K_{i})^2\right)\left(\prod_{s} (\ell_2+ K_{s})^2\right)(\ell_1-\ell_2+R)^2}~~~~\label{IAB}\eea
where ${\cal E} $ is the product of poles involving only external momenta.
%
To proceed, just like the one-loop case \cite{Geyer:2014fka,Baadsgaard:2015hia}, we do the partial fraction using following identity\footnote{The integrand of the type (B-3) in Figure \ref{Fe-1} is given by ${1\over \ell_1^2 (\prod_{i=1}^m (\ell_1+P_i)^2) \ell_1^2 (\ell_1-\ell_2)^2 \ell_2^2}$. The appearance of $(\ell_1^2)^2$ will make the application of partial fraction tricky. We will discuss these contributions later. Similar thing happens to the type (A-1). }
\bea {1\over \prod_{i=1}^n D_i} & = & \sum_{i=1}^n {1\over D_i \prod_{j\neq i} (D_j-D_i)}~~~\label{PF}\eea
and then make  loop momenta shifting.
For the type ${\cal I}_{A}$, the partial fraction and loop momentum shifting will give the standard form ${1\over \ell_1^2\ell_2^2} {1\over \prod P_i}$ with pole $P_i$'s having following   combinations like
$2\ell_i\cdot (K_j-K_i)+ (K_j-K_i)^2=[\ell_i,k_{t_1},...,k_{t_m}]$ (see the notation \eref{notation-1}) if $K_j-K_i=\sum_{i=1}^m k_{t_i}$. These poles are the familiar ones  appearing in the mapping rule \eref{tree-pole} \footnote{It is also worth to notice that it is these contractions $2k_i\cdot k_j$ appearing in the numerator of scattering equations.}
at the tree and one-loop levels. Thus it will be not so surprising that using the same mapping rule reviewed in the section two we can easily read out corresponding expressions given a  CHY-integrand.


However, for the type ${\cal I}_{B}$, things are not so simple because now we have a mixed propagator\footnote{For two-loop planar diagram, there is at most one mixed propagator.} $(\ell_1-\ell_2+R)^2$. When we do the partial fraction of $\ell_1$, should we include the mixed propagator $(\ell_1-\ell_2+R)^2$ in \eref{PF} or not? It is easy to see that if we include the mixed propagator, then we will have terms like ${1\over (\ell_1-\ell_2+R)^2}
{1\over \prod_{i}((\ell_1+K_i)^2-(\ell_1-\ell_2+R)^2)}$. To have the standard ${1\over \ell_1^2\ell_2^2}$ factor in \eref{mloop}, we need to shift $\ell_1=\WH \ell_1+\ell_2-R$. Although it is nothing wrong with this manipulation, the ending pole
$((\ell_1+K_i)^2-(\ell_1-\ell_2+R)^2)$ will become $2\WH\ell_1\cdot(\ell_2-R+K)^2+(\ell_2-R+K)^2$, where although we have linearized the $\WH\ell_1$, the $\ell_2^2$ will appear. The appearance of $\ell_2^2$ will
make the next partial fraction of  $\ell_2$ very complicated. Furthermore, the mapping rule succussed
at the tree and one-loop levels can not cooperate the term $\ell_2^2$. To avoid these troubles,
one possible way  is to
exclude the mixed propagator when we do the partial fraction over both loop momenta $\ell_1$ and $\ell_2$, then we  will arrive the sum of terms like ${1\over \ell_1^2\ell_2^2 (\ell_1-\ell_2+R)^2 \prod_{i}\prod_{j=1,2}[\ell_j, k_{i_1},...,k_{i_m}]}$. Although the linearized poles fit to the mapping rule, the remaining
mixed propagator $(\ell_1-\ell_2+R)^2$ does not.

Is there a frame such that both features mentioned in previous paragraph (i.e., the partial fraction without the mixed  propagator and the applicability of the mapping rule) can be preserved? The answer is yes if we lift the massive loop momenta in $D$-dimension to massless loop momenta in $(D+d)$-dimension as discussed in previous section. As discussed in the paper \cite{Baadsgaard:2015twa}, the procedure of partial fraction can be understood as taking the residue of poles containing dimensional deformed  loop momenta. More explicitly, let us deform the loop momenta from $D$-dimension to $(D+d)$-dimension $\ell_i\to \W\ell_i=\ell_i+\eta_i$. Under this deformation, we have
\bea (\ell_i+P)^2 \to (\W\ell_i+P)^2 =(\ell_i+P)^2-\eta_i^2\equiv (\ell_i+P)^2+z_i\eea
as well as
\bea (\ell_1-\ell_2+R)^2\to (\W\ell_1-\W\ell_2+R)^2=(\ell_1-\ell_2+R)^2-(\eta_1-\eta_2)^2\eea
As long as $d\geq 2$, we have the freedom to make different choices for $(\eta_1-\eta_2)^2$ while keeping $-\eta_i^2=z_i$ invariant. In \cite{Baadsgaard:2015twa}, the choice made is that $-(\eta_1-\eta_2)^2=z_3$ as a new independent variable,
while for current paper, we will make the choice $-(\eta_1-\eta_2)^2=0$. This can be achieved, for example, taking
\bea \eta_1=(x+iy ,x-iy),~~~\eta_2=(i x+z,i x-z)\eea
with
\bea \left\{x= -\frac{i (z_1-z_2)}{\sqrt{(8+8 i) z_1-(8-8 i)
   z_2}},y= -\frac{i ((2+i) z_1+i z_2)}{\sqrt{(8-8 i)
   z_2-(8+8 i) z_1}},z= \frac{z_1+(1+2 i) z_2}{2 \sqrt{2}
   \sqrt{(-1-i) (z_1+i z_2)}}\right\}\eea
thus $(\eta_1-\eta_2)^2=0$ for all $z_1, z_2$. Under this choice
\bea {\cal I}_{B}(z_1,z_2) & = & {1\over  {\cal E} (\ell_1-\ell_2+R)^2}T_1(z_1) T_2(z_2),\nn
T_1(z_1)&=& {1\over \prod_{i} ((\ell_1+ K_{i})^2+z_1)},~~~~~T_2(z_2)={1\over \prod_{s} ((\ell_2+ K_{s})^2+z_2)}~.\eea
It is easy to see that using the contour integration $\oint {dz_1\over z_1} T_1(z_1)$ we can write down\footnote{For this simple case, there is no residue at  $z_1=\infty$.}
\bea T_1(z_1=0) & = & \sum_i {1\over (\ell_1+ K_{i})^2} {1\over \prod_{j\neq i} ((\ell_1+ K_{j})^2-(\ell_1+ K_{i})^2)}\nn
& \sim & \sum_i {1\over \ell_1^2} {1\over \prod_{j\neq i} (2\ell_1\cdot (K_j-K_i)+(K_j-K_i)^2)} \eea
where at the second line we have shifted the loop momentum to become standard form, which is legitimate under the proper regularization of loop integration (such as the dimensional regularization). Similar expression for $T_2(z_2=0)$ can be written down too. Above manipulation is nothing, but the partial fraction where the mixed propagator
$(\ell_1-\ell_2+R)^2$ is not touched, which is exactly what  we want.
Furthermore,  locations of poles impose   on-shell conditions $\W\ell_i^2=0$, $i=1,2$,  thus
the mixed propagator can be written as
\bea (\ell_1-\ell_2+R)^2=(\W\ell_1-\W\ell_2+R)^2= -2\W\ell_1\cdot \W\ell_2 +R^2+2 R\cdot (\W\ell_1-\W\ell_2)=[\W\ell_1,-\W\ell_2, r_1,...,r_m],~~~R=\sum_i k_{r_i} \eea
which is exactly the right pole structure given in the mapping rule \eref{tree-pole}.

Overall, under this new perspective, the two-loop planar integrand can be written as the sum of following terms\footnote{Again the form \eref{IA-split} can not contain contributions from the reducible two-loop diagrams (see type  (B-3) in Figure \ref{Fe-1}), for which we will discuss separately.}
\bea  {1\over \ell_1^2\ell_2^2}
\left\{ \sum_{i,s} {1\over \W{\cal E}\prod_{j\neq i}[\W\ell_1,K_j-K_i]\prod_{t\neq s}[\W\ell_2,K_t-K_s]^2}\right\}|_{\W\ell_1^2=\W\ell_2^2=0}~~~~\label{IA-split}\eea
where $\W{\cal E}$ is ${\cal E}$ for ${\cal I}_A$ and ${\cal E}[\W\ell_1,-\W\ell_2,R]$ for ${\cal I}_B$. From \eref{IA-split} it is clear that the calculation of two-loop integrands is reduced to the
calculation of the part inside the curly bracket. What is the physical picture of these terms?
The on-shell conditions $\W\ell_1^2=\W\ell_2^2=0$ mean that we have cut two loop momenta to on-shell, thus  two-loop diagrams are open up to become tree diagrams with $4$ extra legs with momenta $-\W\ell_1, \W\ell_1, -\W\ell_2, \W\ell_2$. However, as we have discussed before, not all tree diagrams with $(n+4)$-legs can be obtained by this way, especially these coming from one-loop and two tadpoles and one-loop and two-loop massless bubbles (see Figure \ref{Fe-1}). We will discuss this problem in next subsection.

\subsection{Special Feynman diagrams}

\begin{figure}
  \includegraphics[width=15cm]{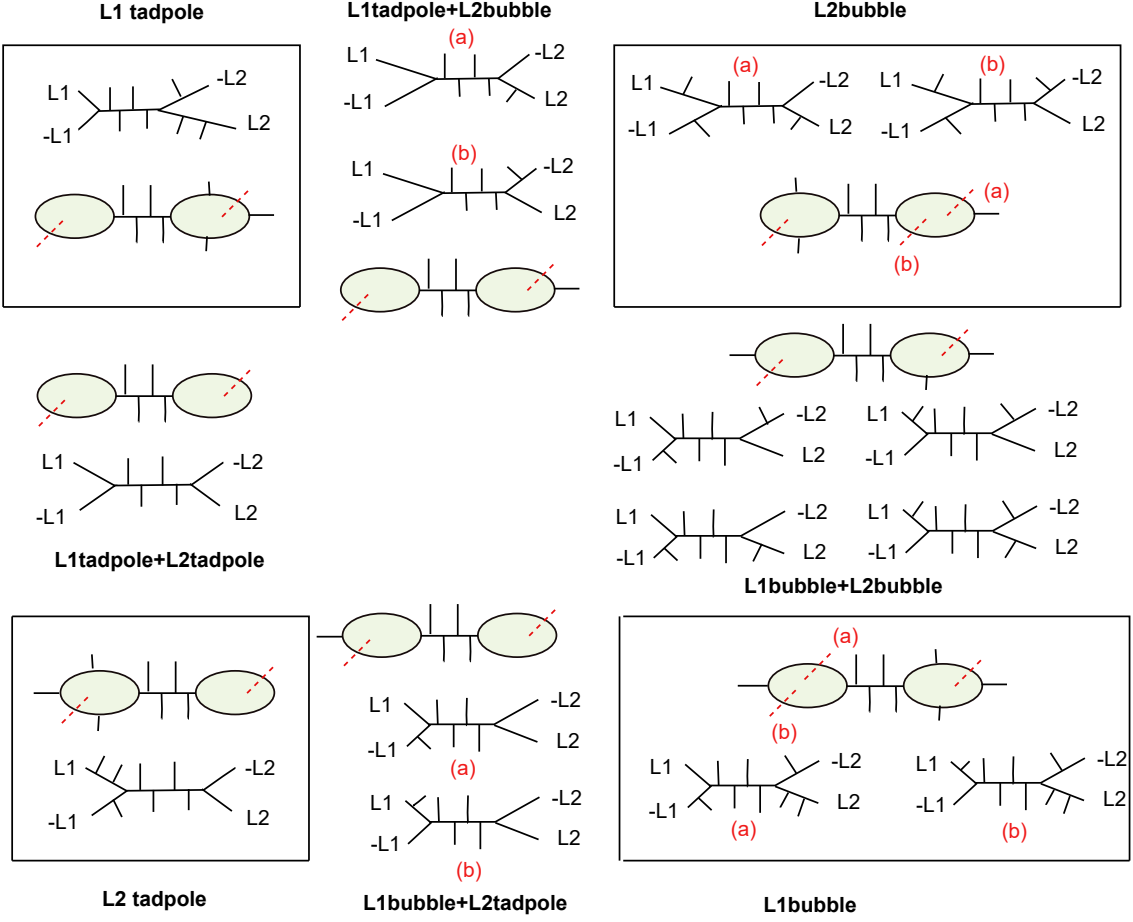}\\
 \caption{Singular contributions from the one-loop tadpoles and one-loop massless bubbles. At the four corners, we have four general cases. For example, the corner "L1-tadpole" means that the left one-loop is tadpole while the right one-loop can be general. Each pair of nearby corners has an intersection. For example, between the corner "L1-tadpole" and the corner "L2-bubble" we will have the diagram
 where the left one-loop is tadpole and the right one-loop is massless bubble. For each loop diagram, we have also drawn the corresponding tree diagrams after the cut. These pictures will be very useful when we discuss how to write down the CHY-integrand.
  }\label{Two-oneloop}
\end{figure}

In this subsection, we focus on these special diagrams given in Figure \ref{Fe-1}. Among them,
tadpoles and massless bubbles are singular, thus we should remove  corresponding contributions of these tree diagrams,
obtained after cutting two internal propagators from these singular two-loop diagrams, from \eref{IA-split}. To be able to do so, we need to have a better understanding of these tree diagrams.

Let us start from the one-loop tadpoles (A-1) and one-loop massless bubbles (A-2) in Figure \ref{Fe-1}.
Depending on if the left sub-oneloop or right sub-oneloop are tadpoles or massless bubbles, we have four different
combinations, which are given by four boxed corners in Figure \ref{Two-oneloop}. For  the upper-left corner, it is the left sub-oneloop having tadpole structure while the right sub-oneloop can have  arbitrary structure. After cutting two loop propagators, we get  corresponding tree diagrams with $(n+4)$-legs. However, all these diagrams have a common feature: all of them contain the pole $s_{(-\ell_1)\ell_1}$. We will call it the "{\bf signature}" of tadpole structure.
For the lower-right corner, the left sub-oneloop has the massless bubble structure while the right sub-oneloop can have  arbitrary structure. After cutting two loop propagators, we get corresponding tree diagrams having following
"signature" of massless bubble structure: either having pole $s_{(-\ell_1)p}s_{(-\ell_1)\ell_1p}$ or having pole
$s_{\ell_1 p}s_{(-\ell_1)\ell_1p}$ with $p$, the massless leg. Similar analysis can be done for  the upper-right corner where the right sub-oneloop has the massless bubble structure and the lower-left corner where the right sub-oneloop has the tadpole structure.

Above four corners have included all singular behaviors for sub-oneloop structure in two-loop diagrams. However, they
are not completely separate from each other. For example, we can have the special case where both left and right sub-oneloops have the tadpole structure. This has been given in the middle between the upper-left corner and the lower-left corner in Figure \ref{Two-oneloop}. The signature of corresponding tree diagrams is the appearance of  poles $s_{(-\ell_1)\ell_1}$ and $s_{(-\ell_2)\ell_2}$ at same time. Similar phenomena happens for each pair of corners near each other and we have listed them all in Figure \ref{Two-oneloop}.

\begin{figure}
 \includegraphics[width=15cm]{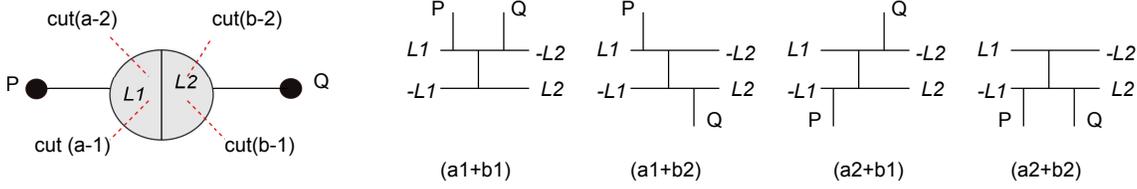}\\
  \caption{ The two-loop massless bubble and its corresponding tree diagrams after various combinations of  cuts. }\label{2loop-exclude-bubble}
\end{figure}

Having understood the one-loop tadpole and massless bubble singularities, we move to the two-loop tadpole and massless bubble singularities. For two-loop massless bubble given in Figure \ref{2loop-exclude-bubble}, depending on  different combinations of cuts, we have four different tree diagrams. Among these four cases, the forward pairs $(-\ell_1, \ell_1)$ and $(-\ell_2, \ell_2)$ are next to each other only in two cases. The signature of these four cases are
$s_{(-\ell_1)\ell_2} s_{P\ell_1} s_{Q(-\ell_2)}$, $s_{Q(-\ell_1)\ell_2} s_{P\ell_1} s_{Q\ell_2}$, $s_{p(-\ell_1)\ell_2} s_{P(-\ell_1)} s_{Q(-\ell_2)}$ and $s_{\ell_1(-\ell_2)} s_{P(-\ell_1)} s_{Q\ell_2}$ with $P+Q=0$. Furthermore,  depending upon if $P$ or $Q$ are massless particle, we need to add another pole
$s_P$ or $s_Q$.

\begin{figure}
 \includegraphics[width=15cm]{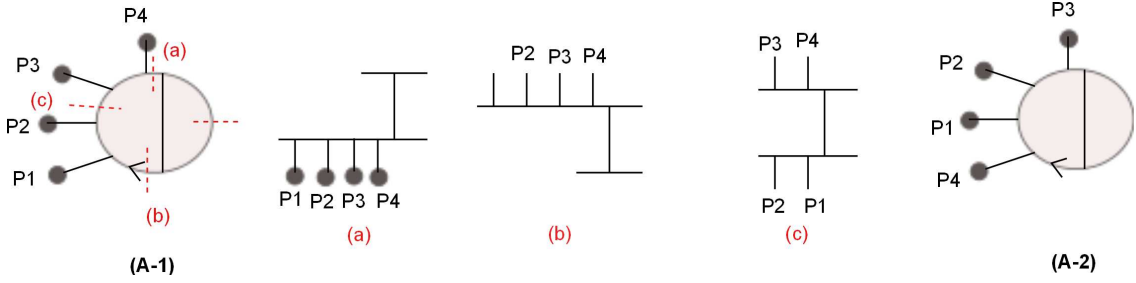}\\
  \caption{ The reducible two-loop diagram (A-1) and its corresponding tree diagrams (a), (b), (c) after cuts. The two-loop diagram (A-2) are obtained from (A-1) by a cyclic permutation. (A-1) and (A-2) give different contributions and we should sum over all cyclic permutations, plus the symmetrization between $\ell_1, \ell_2$.  }\label{Reducible-2loop}
\end{figure}

To discuss the two-loop tadpole, let us start with the (B-3) in Figure \ref{Fe-1}. Since all external legs are attached to one side, the integrand will have the form (see the (A-1) of Figure \ref{Reducible-2loop})
\bea {1\over \ell_1^2 (\prod_{i=1}^m (\ell_1+P_i)) \ell_1^2 (\ell_1-\ell_2)^2 \ell_2^2}~,~~\label{Red-2loop}\eea
where the appearance of $(\ell_1^2)^2$ makes the naive application of  partial fraction to \eref{IAB} problematically. Thus we should not expect to reduce these contributions to the form
\eref{IA-split}. Then how to deal with them? One hint is to rearrange \eref{Red-2loop} as
\bea {1\over \ell_1^2 (\ell_1-\ell_2)^2 \ell_2^2}\left\{{1\over \ell_1^2 (\prod_{i=1}^m (\ell_1+P_i)) } \right\}~,~~~\label{red-one}\eea
then the part inside the bracket is nothing, but the familiar one-loop integrand. However, there is
one subtle point regarding to the choice of loop momentum $\ell_1$. With the convention given in (A-1) and (A-2) of Figure \ref{Reducible-2loop}, it is easy to see that although when rewriting  to the form  \eref{red-one}, both produce  one-loop integrands with the same color ordering, these two one-loop integrands are not same since they have different conventions of loop momentum $\ell_1$ inherited
from two-loop diagrams (although they are related by loop momentum shifting). With above understanding,  we can write two-loop integrands coming from the type (B-3) in Figure \ref{Fe-1} as
\bea {\cal I}^{2-loop}_{B_3}= \left\{{1\over \ell_1^2 (\ell_1-\ell_2)^2 \ell_2^2}\left\{  {\cal I }^{1-loop}
(1,2,...,n,\ell_1)+{\rm cyclic~permu}\{1,2,...,n\} \right\}\right\} + \{\ell_1\leftrightarrow \ell_2\}~~~~\label{red-one-1}\eea

Now we give some explanations for \eref{red-one-1}. First, in each one-loop diagram of ${\cal I }^{1-loop}(1,2,...,n,\ell_1)$, the loop propagator at  the right of the  vertex where leg $1$ has connected  to is defined to be $\ell_1$. Secondly, the two-loop tadpole diagram (B-1) in  Figure \ref{Fe-1} is reduced to
the one-loop tadpole diagram, thus if we exclude these contributions from tadpole diagrams in  ${\cal I }^{1-loop}(1,2,...,n,\ell_1)$, we have excluded the two-loop tadpole contributions. Thirdly, since
we have reduced the calculation of ${\cal I}^{2-loop}_{B_3}$ to one-loop case, we can take them as known data. Thus when we try to find the CHY-construction of two-loop integrands in \eref{IA-split},
we can exclude ${\cal I}^{2-loop}_{B_3}$ part. The complete planar two-loop integrand will be the sum of the result \eref{IA-split} and the result \eref{red-one-1}.
This will be the strategy we follow in the later part of the paper although in the subsection 4.4 we do give a CHY-construction of the ${\cal I}^{2-loop}_{B_3}$ part as the soft limit of a corresponding theory with $(n+5)$-points.

\subsection{The construction of  CHY-integrand}

\begin{figure}
  \includegraphics[width=15cm]{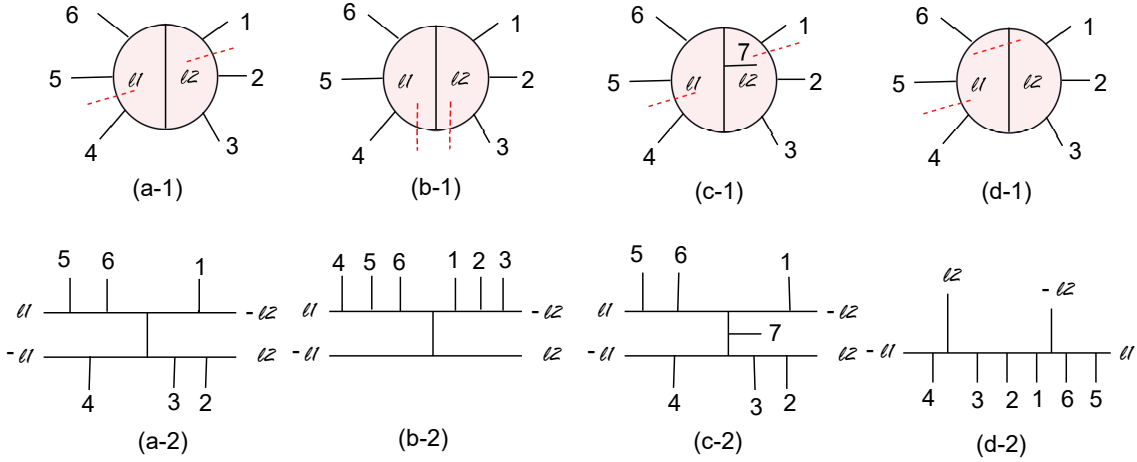}\\
  \caption{Planar two-loop diagrams with two cuts and their corresponding color ordered tree diagrams. In (a),(b), $-\ell_i, \ell_i$ are next to each other while in (c), (d) it is not true anymore.}\label{Fe-cut}
\end{figure}

Having reduced the problem of finding loop integrands to tree diagrams  in \eref{IA-split} (after excluding the ${\cal I}^{2-loop}_{B_3}$ part), we need to determine which tree diagrams we should consider. Since these tree diagrams are  obtained from planar two-loop diagrams by cutting two internal propagators, we can get general picture by speculating the Figure \ref{Fe-cut}. By checking different combinations of two cuts, such as these in (a-1) and (b-1), we can see that in the resulted color ordered tree diagrams (such as these in (a-2) and (b-2)), $-\ell_1$ is always next to $\ell_1$ (similar for the pair $-\ell_2, \ell_2$). This pattern does not hold anymore for non-planar two-loop diagrams (see (c-2)) or cutting along the mixed propagator (see (d-2)). Using this observation, we conclude that the resulted tree diagrams are these obtained
from the original  color ordered $n$-point tree diagrams after inserting two pairs $(-\ell_i,\ell_i)$ consistently
to all possible locations. More explicitly, we will have following two types of ordered diagrams with $(n+4)$-points:
\begin{itemize}

\item Type (I): there are $2n$ of them having  following ordering
\bea {\cal O}_{j}\equiv\{1,...,j,(-\W\ell_1,\W\ell_1), (-\W\ell_2,\W\ell_2),j+1,...,n\}~~~~\label{T1-order}\eea
where $j=1,2,...,n$ (plus also the symmetrization of $\W\ell_1\leftrightarrow \W\ell_2$)\footnote{The symmetrization is necessary since there is no canonical definition of two loops.}.

\item Type (II): there are $n(n-1)$ of them having following ordering
\bea {\cal O}_{jk}\equiv\{1,...,j,(-\ell_1,\ell_1),j+1,...,k, (-\ell_2,\ell_2),k+1,...,n\}~~~~\label{T2-order}\eea
with $1\leq j<k\leq n$ (plus also the symmetrization of $\W\ell_1\leftrightarrow \W\ell_2$).

\end{itemize}
Having found related color ordered tree amplitudes, we know immediately that the part inside the bracket of \eref{IA-split} is the sum of these color ordered tree level amplitudes of type (I) and (II), after removing possible forward singularities and the ${\cal I}^{2-loop}_{B_3}$ part contained in them. Thus the wanted CHY-integrand ${\cal I}^{CHY}$ in \eref{mloop-integral} should produce these contributions. To find
it, we need to use the mapping rule established in \cite{Baadsgaard:2015voa, Baadsgaard:2015ifa,Baadsgaard:2015hia}. Now we discuss one by one.

\subsubsection{The CHY-integrand for ordering ${\cal O}_{jk}$}

Having above general discussions, now we determine the CHY-integrand for each ordering in \eref{T1-order} and \eref{T2-order}. Let us start with the ordering ${\cal O}_{jk}$. With this ordering, the full tree-level amplitude is given by following CHY-integrand
\bea T_{jk}= {1\over z_{12}^2... z_{j(-\ell_1)}^2 z_{(-\ell_1)\ell_1}^2 z_{\ell_1 (j+1)}^2...z_{k(-\ell_2)}^2 z_{(-\ell_2)\ell_2}^2 z_{\ell_2 (k+1)}^2... z_{n1}^2}~.~~~\label{Tjk-0}\eea
Now we consider various forward limits, which can be produced in this ordering. For this purpose, the Figure \ref{Two-oneloop}, the Figure \ref{2loop-exclude-bubble} and the Figure \ref{Reducible-2loop} are very useful.
From these Figures, we see that this ordering can contain following singularities:
\begin{itemize}

\item First it can contain the $\ell_1$-tadpole singularity, i.e., with pole $s_{(-\ell_1)\ell_1}$. These tree diagrams are obtained by the CHY-integrand (please see the Appendix A for full explanations)
    \bea T_{jk;t_1}={1\over z_{12}^2... z_{j(-\ell_1)} \underline{z_{j\ell_1}} z_{(-\ell_1)\ell_1}^2  \underline{z_{(-\ell_1)(j+1)}} z_{\ell_1 (j+1)}...z_{k(-\ell_2)}^2 z_{(-\ell_2)\ell_2}^2 z_{\ell_2 (k+1)}^2... z_{n1}^2}~~~~\label{Tjk-t1-0} \eea
    where we use the $t_1$ to denote the $\ell_1$-tadpole singularity and the underline to emphasize the changed factor. We can write \eref{Tjk-t1-0} to more compact way by using the rule \eref{single-pole-rep}
    \bea T_{jk;t_1}=T_{ij} {\cal P}[j,-\ell_1, \ell_1, j+1]~.~~~\label{Tjk-t1} \eea

\item Secondly it contains  massless $\ell_1$-bubble singularities, i.e., these given by pole structures
    $s_{j(-\ell_1)} s_{j(-\ell_1)\ell_1}$ or $s_{\ell_1 (j+1)}s_{(-\ell_1)\ell_1 (j+1)}$. Using the rule \eref{single-pole-rep} we can write down the corresponding CHY-integrands as
    \bea T_{jk;b_1j}&= & T_{jk} {\cal P}[j-1,j,-\ell_1,\ell_1] {\cal P}[j-1,j,\ell_1, j+1] ~~~~\label{Tjk-b1j} \eea
    and
    \bea T_{jk;b_1(j+1)} & = &  T_{jk} {\cal P}[-\ell_1,\ell_1,j+1, j+2] {\cal P}[j,-\ell_1, j+1,j+2] ~~~~\label{Tjk-b1j+1}\eea
    where we use $b_1$ for massless bubble involving the $\ell_1$ and $j$ to denote the massless bubble of $j$-th leg.

    We want to emphasize one thing: above three singularities are not compatible, i.e., they can not appear at same time in a given tree diagram. Thus when we subtract their contributions, we should subtract all of them.

\item Similar considerations can be done for the $\ell_2$ part and we get following three CHY-integrands
    \bea T_{jk;t_2}& = & T_{ij} {\cal P}[k,-\ell_2, \ell_2, k+1]~~~~\label{Tjk-t2}\\
     T_{jk;b_2 k} & = & T_{jk} {\cal P}[k-1,k,-\ell_2,\ell_2] {\cal P}[k-1,k,\ell_2, k+1] ~~~~\label{Tjk-b2k} \\
      T_{jk;b_2(k+1)}& = & T_{jk} {\cal P}[-\ell_2,\ell_2,k+1, k+2] {\cal P}[k,-\ell_2, k+1,k+2]~~~~\label{Tjk-b2k+1}\eea
    corresponding to one-loop tadpoles and one-loop massless bubbles of $\ell_2$-loop.

\item Now coming to an important observation: the one-loop tadpole and one-loop massless bubble singularities
of $\ell_1$ are (almost) compatible with  the one-loop tadpole and one-loop massless bubble singularities
of $\ell_2$. Thus we will have following nine CHY-integrands to describe tree diagrams having both kinds of singularities. They are:
\bea  T_{jk;t_1, t_2}& = & T_{ij}{\cal P}[j,-\ell_1, \ell_1, j+1] {\cal P}[k,-\ell_2, \ell_2, k+1]~,~~~\label{Tjk-t1-t2}\\
     T_{jk;t_1,b_2 k} & = & T_{jk} {\cal P}[j,-\ell_1, \ell_1, j+1]{\cal P}[k-1,k,-\ell_2,\ell_2] {\cal P}[k-1,k,\ell_2, k+1] ~,~~~\label{Tjk-t1-b2k} \\
      T_{jk;t_1,b_2(k+1)}& = & T_{jk}{\cal P}[j,-\ell_1, \ell_1, j+1] {\cal P}[-\ell_2,\ell_2,k+1, k+2] {\cal P}[k,-\ell_2, k+1,k+2]~,~~~\label{Tjk-t1-b2k+1}\eea
and
\bea  T_{jk;b_1 j, t_2}& = & T_{ij}{\cal P}[j-1,j,-\ell_1,\ell_1] {\cal P}[j-1,j,\ell_1, j+1] {\cal P}[k,-\ell_2, \ell_2, k+1]~,~~~\label{Tjk-b1j-t2}\\
     T_{jk;b_1 j,b_2 k} & = & T_{jk}{\cal P}[j-1,j,-\ell_1,\ell_1] {\cal P}[j-1,j,\ell_1, j+1]{\cal P}[k-1,k,-\ell_2,\ell_2] {\cal P}[k-1,k,\ell_2, k+1] ~,~~~\label{Tjk-b1j-b2k} \\
      T_{jk;b_1 j,b_2(k+1)}& = & T_{jk}{\cal P}[j-1,j,-\ell_1,\ell_1] {\cal P}[j-1,j,\ell_1, j+1]{\cal P}[-\ell_2,\ell_2,k+1, k+2] \nn & & {\cal P}[k,-\ell_2, k+1,k+2]~,~~~\label{Tjk-b1j-b2k+1}\eea
and
\bea  T_{jk;b_1(j+1), t_2}& = & T_{ij}{\cal P}[-\ell_1,\ell_1,j+1, j+2] {\cal P}[j,-\ell_1, j+1,j+2] {\cal P}[k,-\ell_2, \ell_2, k+1]~,~~~\label{Tjk-b1j+1-t2}\\
     T_{jk;b_1(j+1),b_2 k} & = & T_{jk} {\cal P}[-\ell_1,\ell_1,j+1, j+2] {\cal P}[j,-\ell_1, j+1,j+2]{\cal P}[k-1,k,-\ell_2,\ell_2] \nn & & {\cal P}[k-1,k,\ell_2, k+1] ~,~~~\label{Tjk-b1j+1-b2k} \\
      T_{jk;b_1(j+1),b_2(k+1)}& = & T_{jk}{\cal P}[-\ell_1,\ell_1,j+1, j+2] {\cal P}[j,-\ell_1, j+1,j+2] {\cal P}[-\ell_2,\ell_2,k+1, k+2] \nn & & {\cal P}[k,-\ell_2, k+1,k+2]~.~~~~~~\label{Tjk-b1j+1-b2k+1}\eea
There is one warning: when there is only one leg between two pairs of loop momenta, among above nine combinations, some combinations can not exist. More explicitly,  when $k=j+1$, the combination
$T_{jk; b_1 (j+1), b_2 k}$ can not exist, while when  $k=n, j=1$ the combination $T_{jk; b_1j, b_2(k+1)}$ can not exist.

The reason to discuss the compatible structure is to not overly subtract the singular part. For example, after we subtract $T_{jk;t_1}$ and $T_{jk;t_2}$ from $T_{jk}$,  the part $T_{jk;t_1, t_2}$ has been subtracted two times, thus we need to add the  $T_{jk;t_1, t_2}$ part to compensate.

\item Having excluded one-loop singularities,  we continue to remove  two-loop massless bubble singularities.
Although a little bit of away from our concern, let us start with the bubble structure (so including the massive bubble). From the Figure \ref{2loop-exclude-bubble}, we see that two-loop bubble structure will have five fixed poles. With the signature $s_{(j+1)...k}s_{\ell_1(j+1)...k} s_{(k+1)...j} s_{\ell_2(k+1)...j} s_{\ell_2(k+1)...j(-\ell_1)}$ the corresponding CHY-integrand is
    \bea T_{jk;2m1} & = & T_{jk} {\cal P}[\ell_1,j+1,k,-\ell_2] {\cal P}[-\ell_1, \ell_1,k,-\ell_2] {\cal P}[\ell_2,k+1,j,-\ell_1]\nn & & {\cal P}[-\ell_2,\ell_2, j,-\ell_1]{\cal P}[-\ell_2, \ell_2, -\ell_1,\ell_1]~~~~~~~~\label{Tjk-2m-1}\eea
    while with the signature $s_{(j+1)...k}s_{(j+1)...k(-\ell_2)} s_{(k+1)...j} s_{(k+1)...j(-\ell_1)} s_{\ell_2(k+1)...j(-\ell_1)}$ the corresponding CHY-integrand is
    \bea T_{jk;2m2} & = & T_{jk}{\cal P}[\ell_1, j+1, k,-\ell_2] {\cal P}[\ell_1, j+1,-\ell_2, \ell_2] {\cal P}[\ell_2, k+1,j,-\ell_1] \nn & & {\cal P}[\ell_2, k+1, -\ell_1, \ell_1]
    {\cal P}[-\ell_2, \ell_2, -\ell_1, \ell_1]~~~~~~~~\label{Tjk-2m-2}\eea
    Above pole structures with five $s$-factors are general. To get the massless bubble, we need  pole $s_{(j+1)...k}= s_{(k+1)...j}$ to be zero. This can happen only when $k=j+1$ or $k=n, j=1$.
   For $k=j+1$, the signature of  $T_{jk;2m1}$ type reduces to $s_{\ell_1(j+1)...k} s_{(k+1)...j} s_{\ell_2(k+1)...j} s_{\ell_2(k+1)...j(-\ell_1)}$ and  the corresponding  CHY-integrand is
    \bea T_{jk;2m1} & = & T_{jk}  {\cal P}[-\ell_1, \ell_1,k,-\ell_2] {\cal P}[\ell_2,k+1,j,-\ell_1] {\cal P}[-\ell_2,\ell_2, j,-\ell_1]{\cal P}[-\ell_2, \ell_2, -\ell_1,\ell_1]~~~~~~~~\label{Tjk-2m-1-k=j+1}\eea
    while the signature of  $T_{jk;2m2}$ type reduces $s_{(j+1)...k(-\ell_2)} s_{(k+1)...j} s_{(k+1)...j(-\ell_1)} s_{\ell_2(k+1)...j(-\ell_1)}$ and the corresponding CHY-integrand is
    \bea T_{jk;2m2} & = & T_{jk}  {\cal P}[\ell_1, j+1,-\ell_2, \ell_2] {\cal P}[\ell_2, k+1,j,-\ell_1]  {\cal P}[\ell_2, k+1, -\ell_1, \ell_1]
    {\cal P}[-\ell_2, \ell_2, -\ell_1, \ell_1]~~~~~~~~\label{Tjk-2m-2-k=j+1}\eea
    These two expressions \eref{Tjk-2m-1-k=j+1} and \eref{Tjk-2m-2-k=j+1} can be obtained from \eref{Tjk-2m-1} and  \eref{Tjk-2m-2} by trivially setting $k=j+1$ since ${\cal P}[\ell_1, j+1, k,-\ell_2]|_{k=j+1}=1$. Similarly for the case $k=n, j=1$, the corresponding CHY-integrands can be also obtained from \eref{Tjk-2m-1} and  \eref{Tjk-2m-2} by trivially setting $k=n, j=1$ since ${\cal P}[\ell_2, k+1,j,-\ell_1]|_{k=n,j==1}=1$.

    These two are not compatible to each other. They are also not compatible with one-loop tadpole and one-loop massless bubble singularities.

\item The last piece we need to exclude is the (B-3) part in Figure \ref{Fe-1}. From Figure \ref{Reducible-2loop}, we see that with the signature   $s_{\ell_1(j+1)...k}  s_{(k+1)...j(-\ell_1)} s_{\ell_2(k+1)...j(-\ell_1)}$ the corresponding CHY-integrand is
    \bea T_{jk;B31} & = & T_{jk}{\cal P}[-\ell_1, \ell_1, k,-\ell_2] {\cal P}[\ell_2, k+1, -\ell_1, \ell_1] {\cal P}[-\ell_2, \ell_2, -\ell_1, \ell_1]~~~~~~~~\label{Tjk-B3-1}\eea
while with the signature $s_{(j+1)...k(-\ell_2)}  s_{\ell_2(k+1)...j} s_{\ell_2(k+1)...j(-\ell_1)}$ the corresponding CHY-integrand is
    \bea T_{jk;B32} & = & T_{jk} {\cal P}[\ell_1,j+1, -\ell_2, \ell_2]
    {\cal P}[-\ell_2, \ell_2, j,-\ell_1]{\cal P}[-\ell_2, \ell_2, -\ell_1, \ell_1]~~~~~~~~\label{Tjk-B3-2}\eea
    These two are not compatible to each other. They are also not compatible with one-loop tadpole, one-loop massless bubble and two-loop tadpole singularities.
\end{itemize}

Having above analysis, now we can write down the wanted CHY-integrand for the ordering ${\cal O}_{jk}$ as
\bea {\cal I}^{CHY}_{{\cal O}_{jk}} & = & T_{jk} -\left( T_{jk; t_1}+ T_{jk; b_1 j}+T_{jk; b_1 (j+1)}
+T_{jk; t_2}+ T_{jk; b_2 k}+T_{jk; b_2 (k+1)}\right)\nn
& & +\left( T_{jk; t_1, t_2}+ T_{jk; t_1, b_2 k}+T_{jk; t_1, b_2(k+1)}+  T_{jk; b_1j, t_2}+ T_{jk;b_1j , b_2 k}+(1-\delta_{k,n}\delta_{j,1})T_{jk; b_1 j, b_2(k+1)}\right. \nn & & \left. +T_{jk; b_1(j+1), t_2}+(1-\delta_{j+1,k}) T_{jk; b_1(j+1), b_2 k}+T_{jk; b_1(j+1), b_2(k+1)}\right)\nn
& &- (\delta_{j+1,k}+\delta_{k,n}\delta_{j,1})\left( T_{jk; 2m1}+T_{jk;2m2}\right)- \left( T_{jk; B31}+ T_{jk;B32}\right)~~~~\label{Ojk-CHY} \eea
where we have inserted delta functions for special cases $k=(j+1)$ or $k=n, j=1$.

Before ending this subsection, there is nice feature worth to mention
 about one-loop massless bubble singularities. It is well known that the integration of one-loop massless bubble is zero under proper regularization (such as dimensional regularization). We can also see it clearly at the integrand level in current setup. For one-loop massless bubble, the integrand is given by ${ N(\ell)\over \ell^2(\ell-p)^2}$. After the partial fraction and momentum shifting we get

\bea  {N(\ell)\over \ell^2 (-2\ell\cdot p)}+
 {N(\ell)\over (\ell-p)^2 (2\ell\cdot p)}\simeq  {N(\ell)\over \ell^2 (-2\ell\cdot p)}+
 {N(\ell+p)\over \ell^2 (2\ell\cdot p)}~.\eea
Thus if $N(\ell)=N(\ell+P)$ (which is true for scalar theory), they cancel each other at the integrand level. It is worth to emphasize that the cancelation happens between two different orderings as
having been observed in \cite{Baadsgaard:2015hia}, i.e., the ordering $\{...,-\ell, \ell, p,...\}$
and the ordering $\{...,p,-\ell,\ell,...\}$. For two-loop massless bubble, we can do similar manipulation
\bea & & {N(\ell_1, \ell_2)\over \ell_1^2 (\ell_1+p)^2 (\ell_1-\ell_2+p)^2 \ell_2^2 (\ell_2-p)^2}\nn
& = & {N(\ell_1, \ell_2)\over (\ell_1-\ell_2+p)^2}\left({1\over \ell_1^2 (2\ell_1\cdot p)}
+{1\over (\ell_1+p)^2 (-2\ell_1\cdot p)} \right)\left({1\over \ell_2^2 (-2\ell_2\cdot p)}
+{1\over (\ell_2-p)^2 (2\ell_2\cdot p)} \right)\nn
& \simeq & { N(\ell_1, \ell_2)\over \ell_1^2 \ell_2^2(\ell_1-\ell_2+p)^2(2\ell_1\cdot p)(-2\ell_2\cdot p)}+ { N(\ell_1, \ell_2+p)\over \ell_1^2 \ell_2^2(\ell_1-\ell_2)^2(2\ell_1\cdot p)(2\ell_2\cdot p)}\nn
& & + {N(\ell_1-p, \ell_2)\over \ell_1^2 \ell_2^2(\ell_1-\ell_2)^2(-2\ell_1\cdot p)(-2\ell_2\cdot p)}
+{N(\ell_1-p, \ell_2+p)\over \ell_1^2 \ell_2^2(\ell_1-\ell_2-p)^2(-2\ell_1\cdot p)(2\ell_2\cdot p)}~.
\eea
Since the different mixed propagators $(\ell_1-\ell_2+p)^2$, $(\ell_1-\ell_2-p)^2$ and $(\ell_1-\ell_2+p)^2$, we see that even $N(\ell_1, \ell_2)=1$, they are not cancel each other at the integrand level. Because the explicit cancelation at the integrand level for one-loop massless bubble after summing over all orderings, we can save
the explicit subtraction in \eref{Ojk-CHY} and simplify it to
\bea {\cal I}^{CHY}_{{\cal O}_{jk}} & = &\left\{ T_{jk} -\left( T_{jk; t_1}
+T_{jk; t_2}\right) + T_{jk; t_1, t_2}\right\}- (\delta_{j+1,k}+\delta_{k,n}\delta_{j,1})\left( T_{jk; 2m1}+T_{jk;2m2}\right)- \left( T_{jk; B31}+ T_{jk;B32}\right)~.~~~~~\label{Ojk-CHY-1} \eea
One can sum up the first four terms to simplify to
\bea T_{jk} \left(1- {z_{j(-\ell_1)} z_{\ell_1 (j+1)}\over z_{j\ell_1} z_{(-\ell_1) (j+1)} } \right)\left(1- {z_{k(-\ell_2)} z_{\ell_2 (k+1)}\over z_{k\ell_2} z_{(-\ell_2) (k+1)} }\right)= T_{jk} {z_{j(j+1)} z_{(-\ell_1)\ell_1 }\over z_{j\ell_1} z_{(-\ell_1) (j+1)} } {z_{k(k+1)} z_{(-\ell_2) \ell_2}\over z_{k\ell_2} z_{(-\ell_2) (k+1)} }~. \eea
Although one can continue to add later four terms, \eref{Ojk-CHY-1} has more clear physical picture.

\subsubsection{The CHY-integrand for ordering ${\cal O}_{j}$}

Having done the ordering ${\cal O}_{jk}$, we consider the ordering ${\cal O}_{j}$. With this ordering, the full tree-level amplitude of $(n+4)$-legs  is given by following CHY-integrand
\bea T_{j}= {1\over z_{12}^2... z_{j(-\ell_1)}^2 z_{(-\ell_1)\ell_1}^2 z_{\ell_1(-\ell_2)}^2 z_{(-\ell_2)\ell_2}^2 z_{\ell_2 (j+1)}^2... z_{n1}^2}~~~~\label{Tj-0}\eea
Now we consider various forward limits, which can be produced in this ordering by checking the Figure \ref{Two-oneloop}, the Figure \ref{2loop-exclude-bubble} and the Figure \ref{Reducible-2loop}:
\begin{itemize}

\item First there are  one-loop tadpole singularities, thus we have the corresponding CHY-integrands
\bea T_{j; t_1} & = & T_j {\cal P}[j,-\ell_1,\ell_1, -\ell_2],~~~~~~
T_{j; t_2} = T_j {\cal P}[\ell_1, -\ell_2, \ell_2, j+1]~~~\label{Tj-t1-t2} \eea
for $\ell_1$-tadpoles and $\ell_2$-tadpoles respectively.

\item Secondly, there are one-loop massless bubbles. With $j$-th leg, there are bubbles with poles $s_{j (-\ell_1)} s_{j(-\ell_1)\ell_1}$ and its corresponding CHY-integrand is
    \bea T_{j;b_1 j} &= & T_j {\cal P}[j-1, j, -\ell_1,\ell_1]{\cal P}[j-1, j, \ell_1, -\ell_2]~.~~\label{Tj-b1j}\eea
With $(j+1)$-th leg, there are bubbles with poles $s_{\ell_2(j+1)} s_{(-\ell_2)\ell_2(j+1)}$ and its corresponding CHY-integrand is
    \bea T_{j;b_2 (j+1)} &= & T_j {\cal P}[-\ell_2, \ell_2, j+1, j+2]{\cal P}[\ell_1, -\ell_2, j+1, j+2]~.~~\label{Tj-b2j+1}\eea

\item Again, because the compatibility we have following four combinations between $\ell_1$ one-loop forward singularities and $\ell_2$ one-loop forward singularities:
\bea T_{j; t_1, t_2} & = & T_j {\cal P}[j,-\ell_1,\ell_1, -\ell_2]{\cal P}[\ell_1, -\ell_2, \ell_2, j+1]~~~~\label{Tj-t1-t2} \\
 T_{j; t_1,b_2(j+1)} & = & T_j {\cal P}[j,-\ell_1,\ell_1, -\ell_2]{\cal P}[-\ell_2, \ell_2, j+1, j+2]{\cal P}[\ell_1, -\ell_2, j+1, j+2],~~~~\label{Tj-t1-b2j+1}\\
T_{j; b_1j,t_2} &= & T_j{\cal P}[j-1, j, -\ell_1,\ell_1]{\cal P}[j-1, j, \ell_1, -\ell_2]{\cal P}[\ell_1, -\ell_2, \ell_2, j+1] ~~~~~\label{Tj-t2-b1j}\\
 T_{j; b_1j ,b_2(j+1)} & = & T_j {\cal P}[j-1, j, -\ell_1,\ell_1]{\cal P}[j-1, j, \ell_1, -\ell_2]{\cal P}[-\ell_2, \ell_2, j+1, j+2]\nn & & {\cal P}[\ell_1, -\ell_2, j+1, j+2],~~~~\label{Tj-b1j-b2j+1}
 \eea

\item Now we discuss  two-loop massless bubbles. Again let us start with two-loop bubble topologies. From Figure \ref{2loop-exclude-bubble} we can see that beside the  pole $s_{\ell_1(-\ell_2)}$, there is a free parameter  $k$ with $k=(j+1),(j+2),...,n,1,...,j-1$. For each $k$, there is one bubble structure with pole $s_{\ell_1(-\ell_2)} s_{(j+1)...k} s_{\ell_2(j+1)...k} s_{(k+1)...j}$ $s_{(k+1)...j(-\ell_1)} $ and the corresponding CHY-integrand is given by
    \bea T_{j; 2m}[k] & = & T_{j} {\cal P}[-\ell_1,\ell_1, -\ell_2, \ell_2] {\cal P}[\ell_2, j+1, k,k+1] {\cal P}[-\ell_2, \ell_2, k,k+1] \nn & &  {\cal P}[k,k+1, j, -\ell_1]{\cal P}[k,k+1, -\ell_1,\ell_1]~.~~~\label{Tj-2m}\eea
    Again, for general $k$, they are massive bubbles. Only when $k=j+1$ or $k=j-1$ we get the massless bubbles. The CHY-integrands of both special cases can be trivially  obtained from \eref{Tj-2m} by setting $k=j+1$ or $k=j-1$. For $k=j+1$, the factor ${\cal P}[\ell_2, j+1, k,k+1]=1$ while for $k=j-1$, the factor ${\cal P}[k,k+1, j, -\ell_1]=1$.


\item The last piece we need to exclude is the (B-3) part in Figure \ref{Fe-1}. From Figure \ref{Reducible-2loop}, we see that with the pole $s_{\ell_1(-\ell_2)} s_{\ell_2(j+1)...j}$, the CHY-integrand is given by
    \bea T_{j;B31} & = & T_j {\cal P}[-\ell_1, \ell_1, -\ell_2, \ell_2]{\cal P}[-\ell_2, \ell_2,j,-\ell_1] ~,~~~~\label{Tj-b31}\eea
    while with the pole $s_{\ell_1(-\ell_2)} s_{(j+1)...j(-\ell_1)}$, the CHY-integrand is given by
    \bea T_{j;B32} & = & T_j {\cal P}[-\ell_1, \ell_1, -\ell_2, \ell_2] {\cal P}[\ell_2, j+1, -\ell_1, \ell_1]~.~~~~\label{Tj-b31}\eea

\end{itemize}
Having above analysis,  we can write down the CHY-integrand for the ordering ${\cal O}_{j}$ as
\bea {\cal I}^{CHY}_{{\cal O}_j} & = & T_j - \left( T_{j;t_1}+T_{j;t_2}+T_{j; b_1 j}+T_{j; b_2(j+1)} \right)
+\left( T_{j;t_1,t_2}+ T_{j; t_1, b_2(j+1)}+T_{j;  b_1 j,t_2}+T_{j; b_1 j, b_2(j+1)}\right)\nn
& & -  T_{j; 2m}[j+1]-T_{j; 2m}[j-1]- \left( T_{j;B31}+T_{j;B32}\right)~~~~\label{Oj-CHY} \eea
Again we can forget  one-loop massless bubbles to simplify the expression, although we prefer the more complicated one \eref{Oj-CHY}.

\subsection{The CHY-construction of reducible  two-loop diagrams}

As mentioned in the subsection 4.2, for two-loop diagrams, there are  special two-loop diagrams (called the "reducible two-loop" diagrams), which will cause some troubles when we apply the partial fraction. After careful analysis, we have reduced the problem to the one-loop case in \eref{red-one} and \eref{red-one-1}. Although as we have mentioned, we will treat this part as known data, tn this subsection, we will try to give a direct CHY-construction of these reducible two-loop diagrams at the two-loop level.

\begin{figure}
  \includegraphics[width=15cm]{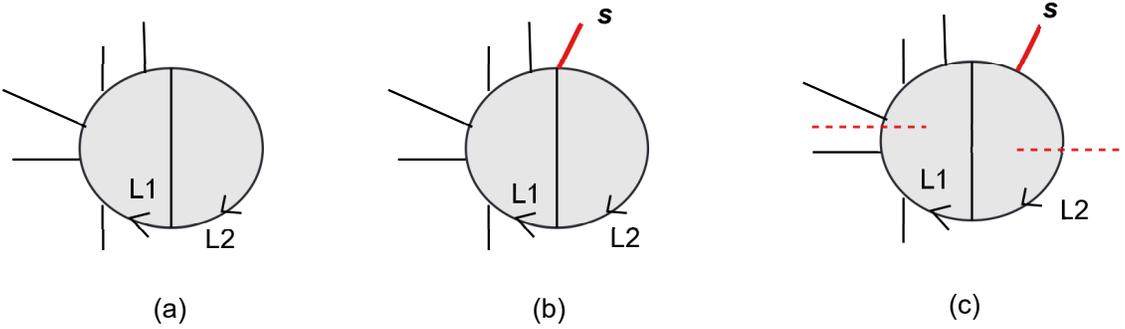}\\
  \caption{ (a) The reducible  two-loop diagrams; (b) After adding a particle at the vertex. However, this vertex has four legs. (c) Moving the added particle to $\ell_2$-loop to make the vertex cubic. Furthermore, we have illustrated possible cuts for this cubic diagram.
  } \label{trouble-2loop}
\end{figure}

Let us recall the general expressions for reducible two-loop diagrams. From (a) of Figure \ref{trouble-2loop} we can read out
\bea {1\over \ell_1^2 (\prod_{i=1}^m (\ell_1+P_i)) \ell_1^2 (\ell_1-\ell_2)^2 \ell_2^2}~~~\label{Old-exp}\eea
under our choice of loop momenta. Now from these $n$ external momenta $k_i$ satisfying $\sum_{i=1}^n k_i=0$, we try to
 construct $(n+1)$ massless momenta by following way. Picking up, for example, $k_n$ and a massless momentum $k_s$ such that $k_n\cdot k_s=0$, then the  $(n+1)$ massless momenta can be arranged to be $\{k_1,...,k_{n-1}, k_{n}-t k_s, t k_s\}$. Using this construction, each diagram (a) in Figure \ref{trouble-2loop} will have a corresponding diagram (b) in Figure \ref{trouble-2loop} with the expression
\bea {1\over \ell_1^2 (\prod_{i=1}^m (\ell_1+P_i)^2) (\ell_1-t k_s)^2 (\ell_1-\ell_2)^2 \ell_2^2}~~~\label{New-exp}\eea
It is easy to see that under the soft limit $t\to 0$, \eref{New-exp} reduces to \eref{Old-exp}. Although it looks nice, it is not the $\phi^3$ theory since we have one vertex with four legs. We can remedy this by moving the leg $s$ to the $\ell_2$-loop as given by the diagram (c) in Figure \ref{trouble-2loop}. Thus \eref{New-exp} can be written as
\bea (\ell_2-t k_s)^2\times {1\over \ell_1^2 (\prod_{i=1}^m (\ell_1+P_i)^2) (\ell_1-t k_s)^2 (\ell_1-\ell_2)^2 \ell_2^2(\ell_2-t k_s)^2}~~~\label{New-exp-1}\eea
Now formula \eref{New-exp-1} can have the CHY-construction by the standard procedure, i.e., partial fraction and momentum shifting, thus we arrive
\bea {-2t\W\ell_2\cdot k_s\over \ell_1^2 \ell_2^2} {\cal A}(\pm\W\ell_1, \pm\W\ell_2,1,...,k_n-t k_s, k_s)~~~\label{red-form} \eea
where the ${\cal A}$ is certain tree-level amplitude with $(n+5)$-points, where  $\W\ell_1, \W\ell_2$ are on-shell
momenta in higher dimension.

Having above picture, now we can present the explicit CHY-construction for the term
$ {1\over \ell_1^2 (\ell_1-\ell_2)^2 \ell_2^2}$ $\times {\cal I }^{1-loop}
(1,2,...,n,\ell_1)$ in \eref{red-one-1} as the soft limit $t\to 0$ of \eref{red-form} (other terms in \eref{red-one-1}
can be obtained by cyclic permutations). The ${\cal A}$ is given by the sum over following
orderings of trees:
\bea {\cal O}_j=\{(-\ell_2, \ell_2), 1,..., j, (-\ell_1, \ell_1), j+1,...,n, s \}, ~~~~j=0,1,...,n~~~\label{red-Oj}\eea
where $j=0$ means the pair $(-\ell_1,\ell_1)$ is inserted before the leg $1$.  For the $j$-th ordering ($j\neq 0,n$), the signature of pole is $s_{s(-\ell_2)}  s_{\ell_1(j+1)...n} s_{\ell_2 1...j(-\ell_1)}
s_{1...j(-\ell_1)}$, thus the CHY-integrand is given by
\bea T_j & = & {\cal I}_{j}{\cal P}[n,s,-\ell_2, \ell_2] {\cal P}[-\ell_1, \ell_1, n,s]
{\cal P}[-\ell_2,\ell_2, -\ell_1, \ell_1]{\cal P}[\ell_2,1,-\ell_1, \ell_1]\eea
with
\bea {\cal I}_{j}= {1\over z_{(-\ell_2)\ell_2}^2 z_{\ell_2 1}^2... z_{j(-\ell_1)}^2 z_{(-\ell_1)\ell_1}^2 z_{\ell_1(j+1)}^2... z_{ns}^2 z_{s(-\ell_2)}^2}~.\eea
For the $j=0$, the signature of pole is $s_{s(-\ell_2)}  s_{\ell_1 1...n} s_{\ell_2 (-\ell_1)}$, thus the CHY-integrand is given by
\bea T_{j=0} & = & {\cal I}_{j=0}{\cal P}[n, s, -\ell_2, \ell_2] {\cal P}[-\ell_1, \ell_1, n,s]
{\cal P}[-\ell_2, \ell_2, -\ell_1, \ell_1]~.\eea
For the $j=n$, the signature of pole is $s_{s(-\ell_2)}  s_{\ell_1 s (-\ell_2)} s_{1...n(-\ell_1)}$, thus the CHY-integrand is given by
\bea T_{j=n} & = & {\cal I}_{j=n}{\cal P}[\ell_1, s,-\ell_2, \ell_2] {\cal P}[-\ell_1, \ell_1, -\ell_2, \ell_2]{\cal P}[\ell_2,1,-\ell_1,\ell_1]~.\eea
Next we need to subtract forward singularities related to above orderings:
\begin{itemize}

\item The tadpole structure can appear when $j=0$ or $j=n$. For $j=0$ the signature will be $s_{s(-\ell_2)}  s_{\ell_1 1...n} s_{\ell_2 (-\ell_1)}$ multiplying by a further factor $s_{1...n}$, thus we have
    \bea T_{j=0;t}= T_{j=0}{\cal P}[\ell_1, 1,n,s]~. \eea
    For $j=n$ the signature will be $s_{s(-\ell_2)}  s_{\ell_1 s (-\ell_2)} s_{1...n(-\ell_1)}$ multiplying by a further factor $s_{1...n}$, thus we have
    \bea T_{j=n;t}= T_{j=n}{\cal P}[\ell_2,1,n,-\ell_1]~.\eea

\item The massless bubble structure can appear when $j=0, 1, n-1, n$. For $j=0$ the signature will be $s_{s(-\ell_2)}  s_{\ell_1 1...n} s_{\ell_2 (-\ell_1)}$ multiplying by either the factor $s_{\ell_1 1} s_{2..n}$ or $s_{1...(n-1)} s_{\ell_1 1...(n-1)}$, thus the corresponding CHY-integrands are
    \bea T_{j=0; b1} & = & T_{j=0}{\cal P}[-\ell_1, \ell_1, 1,2] {\cal P}[1,2,n,s]~,\nn
    T_{j=0; b2} & = & T_{j=0}{\cal P}[-\ell_1, \ell_1, n-1,n]{\cal P}[\ell_1, 1, n-1, n]~.\eea
 For $j=n$ the signature will be $s_{s(-\ell_2)}  s_{\ell_1 s (-\ell_2)} s_{1...n(-\ell_1)}$ multiplying by either the factor $s_{(-\ell_1) n} s_{1...(n-1)} $ or $s_{2...n} s_{2...n(-\ell_1)}$, thus we have
    \bea T_{j=n;b1}& = &  T_{j=n}{\cal P}[n-1,n,-\ell_1, \ell_1] {\cal P}[\ell_2, 1,n-1,n]~,\nn
    T_{j=n;b2}& = &  T_{j=n}{\cal P}[1,2,n,-\ell_1] {\cal P}[1,2,-\ell_1, \ell_1]~.\eea
    For $j=1$, the signature will be the one of $T_{j=1}$ multiplying by a further factor $s_{2...n}$, thus the CHY-integrand is
    \bea T_{j=1;b} & = & T_{j=1}{\cal P}[\ell_1, 2,n,s]~.\eea
    For $j=n-1$, the signature will be the one of $T_{j=n-1}$ multiplying by a further factor $s_{1...(n-1)}$, thus the CHY-integrand is
    \bea T_{j=n-1;b} & = & T_{j=n-1}{\cal P}[\ell_2,1,n-1, -\ell_1]~.\eea
\end{itemize}
Putting all together, we finally arrive following CHY-integrand:
\bea {\cal I}_{\cal A}(1,2...,n) & = & \sum_{j=0}^n T_j -( T_{j=0; t}+ T_{j=n; t})
-( T_{j=0; b1}+T_{j=0;b2}+ T_{j=n; b1} T_{j=n; n2})-(T_{j=1;b}+T_{j=n-1;b})~~~~~~~\eea
Before ending this subsection, we want to remark that although we have provided a solution using the soft limit,
more direct treatment is still preferred, but now we need to understand how to construct CHY-integrands with double poles at the tree-level. This will be an interesting thing to investigate.

\section{Counting}

Having reduced the two-loop problem to tree level (i.e., the loop scattering equations and the loop CHY-integrands)
using the point of view of  dimensional reduction, the checking of the proposal for two-loop becomes the checking of
corresponding tree one. Since the later one has been extensively checked, both numerically and analytically (especially the powerful mapping rule), our proposal should be right. In this section, we will give a further evidence to
support our claim by comparing the number of terms, produced by directly Feynman diagrams or by CHY-formula. In this section, contributions coming from  reducible two-loop diagrams will be excluded.

\subsection{Counting from CHY-formula}

Since the whole result is obtained by summing over $2n$ orderings of ${\cal O}_j$ type and $n(n-1)$ orderings of ${\cal O}_{jk}$ type, we count terms from these two types one by one.

~\\{\bf The ${\cal O}_{j}$ type:} Let us start with the formula \eref{Oj-CHY}. The first term gives the full tree-level amplitude of $(n+4)$-points, so it gives $C(n+4)$ terms. For the $T_{j;t_1}$, since all tree diagrams have the pole $s_{(-\ell_1)\ell_2}$, these two legs have been effectively grouped to become one leg, thus all these diagrams become the tree-level diagrams of the $(n+3)$-points, so it gives $C(n+3)$ terms. For the $T_{j;B31}$  and $T_{j;B32}$, we see that they are effectively tree-level amplitudes of $(n+2)$-points.   Similar arguments give
\bea {\cal N}[T_j]& = & C(n+4),~~~ {\cal N}[ T_{j;t_1}]={\cal N}[T_{j;t_2}]=C(n+3),~~~~{\cal N}[T_{j; b_1 j}]={\cal N}[T_{j; b_2(j+1)}]=C(n+2),~~~\nn
{\cal N}[T_{j;t_1,t_2}] & = &C(n+2),~~~~~ {\cal N}[ T_{j; t_1, b_2(j+1)}]={\cal N}[T_{j;  b_1 j,t_2}]=C(n+1),\nn
{\cal N}[ T_{j; b_1 j, b_2(j+1)}]&=& C(n),~~~~{\cal N}[T_{j;B31}]= {\cal N}[T_{j;B32}]=C(n+2)~~~~~~\eea
For general $T_{j;2m}[k]$, the counting  is a little bit complicated. With a given $k$, we have two tree diagrams: one with $2+(k-j)$-legs and one with $n-(k-j)+2$-legs. Furthermore, external legs in each group will combine together before meeting $\ell_i$ (i.e., the tree-structure of $1+(k-j)$-points  and $n-(k-j)+1$-points), thus we will have the counting $C(k-j+1) C(n-(k-j)+1)$. However, for massless bubbles, we just need to consider the case $k-j=1$ or $n-(k-j)=1$ and both cases give $C(n)$ terms.

%
%
Putting all together, we finally arrive
\bea {\cal N}[{\cal O}_j]
& = &C(n+4)-2C(n+3)-3C(n+2)+2C(n+1)-C(n)~.~~~\label{n-Oj}\eea
This expression does not depend on the value of $j$ as it should. There are $2n$ of this type, so the final number of terms coming from this type should be
\bea {\cal N}_{I} = 2n \left\{ C(n+4)-2C(n+3)-3C(n+2)+2C(n+1)-C(n)\right\}~.~~\label{nI}\eea

~\\{\bf The ${\cal O}_{jk}$ type:} For this one, we start with \eref{Ojk-CHY} with $1\leq j<k \leq n$. Using similar arguments we give the counting for each piece:
\bea  {\cal N}[ T_{jk}] & = &  C(n+4),~~~~~~{\cal N}[ T_{jk; t_1}]={\cal N}[T_{jk; t_2}]=C(n+3),~~~{\cal N}[T_{jk; t_1, t_2}]=C(n+2),\nn
{\cal N}[ T_{jk; b_1 j}]& = & {\cal N}[T_{jk; b_1 (j+1)}]={\cal N}[T_{jk; b_2 k}]={\cal N}[T_{jk; b_2 (k+1)}]=C(n+2),\nn
 {\cal N}[T_{jk; t_1, b_2 k}]& = & {\cal N}[T_{jk; t_1, b_2(k+1)}]={\cal N}[T_{jk; b_1(j+1), t_2}]={\cal N}[ T_{jk;b_1j , b_2 k}]=C(n+1),\nn
{\cal N}[T_{jk;b_1j , b_2 k}] & =& {\cal N}[T_{jk; b_1(j+1), b_2(k+1)}]={\cal N}[ T_{jk; b_1(j+1), b_2 k}]={\cal N}[T_{jk; b_1(j+1), b_2(k+1)}]=C(n),\nn
{\cal N}[ T_{jk; 2m1}] & =& {\cal N}[T_{jk;2m2}]= C(k-j+1) C(n-(k-j)+1),\nn
 {\cal N}[ T_{jk; B31}]&= & {\cal N}[T_{jk;B32}]=C(k-j+2)C(n-(k-j)+2)~.~~~\label{n-Ojk-CHY-details} \eea
Thus from \eref{Ojk-CHY} we get
\bea {\cal N}[{\cal Q}_{jk}] & = &C(n+4)-2C(n+3)-3C(n+2)+4C(n+1)+(4-\delta_{j+1,k}-\delta_{k,n} \delta_{j,1})C(n)\nn & & -2 (\delta_{j+1,k}+\delta_{k,n} \delta_{j,1})C(k-j+1) C(n-(k-j)+1)-2 C(k-j+2)C(n-(k-j)+2)~.~~~\label{n-Ojk-gen} \eea
Thus  when $k=j+1$ or $k=n, j=1$, the counting is given by
\bea {\cal N}^s & = &C(n+4)-2C(n+3)-3C(n+2)+4C(n+1)+3 C(n) -2 C(n)-2 C(n+1)\nn
& = & C(n+4)-2C(n+3)-3C(n+2)+2C(n+1)+ C(n)~.~~~\label{n-Ojk-deg}  \eea
For other cases, we have
\bea {\cal N}^g[{\cal Q}_{jk}] & = & C(n+4)-2C(n+3)-3C(n+2)+4C(n+1)+4 C(n)\nn & &  -2 C(k-j+2)C(n-(k-j)+2)~~.~~~~~\label{n-Ojk}  \eea
Now we sum up all pairs of $(j,k)$. For special cases, there are $2n$ of them, so we have
\bea {\cal N}_{II,A} &= &2n {\cal N}^s =2n \left\{C(n+4)-2C(n+3)-3C(n+2)+2C(n+1)+ C(n) \right\}.~~~\label{nII-A}\eea
For other  cases with number $n(n-1)-2n$, we have the sum
\bea {\cal N}_{II,B} & = & 2\left\{\sum_{k=3}^{n-1} {\cal N}^g[{\cal Q}_{j=1,k}]+ \sum_{j=2}^{n-2}\sum_{k=j+2}^n {\cal N}^g[{\cal Q}_{jk}]\right\}.~~~\label{nII-B}\eea

~\\{\bf Summary:} The total number  of terms given by the CHY-formula is
\bea {\cal N}_{CHY} & = & {\cal N}_{I} + {\cal N}_{II,A}+ {\cal N}_{II,B}~.~~~\label{n-CHY}\eea
%

\subsection{Counting from Feynman diagrams}

Now we do the counting using Feynman diagrams given in Figure \ref{Fe-1} directly. Although we will count terms for
Type (A) and Type (B) separately, they do share same one-loop building block as indicated by the red square in Figure \ref{Fe-1} (the $n_L$ part of Type A), thus we need to consider terms contributing from the building block first. To deal with it, it is crucial to recall that after applying the partial fraction to expression  ${1\over \prod (\ell+K_i)^2}$, we will get terms like ${1\over (\ell_1+K_i)^2}\times F_i$ for each $K_i$. Now we count terms with the same propagator ${1\over (\ell_1+K_i)^2}$. Since the partial fraction has the physical picture as cutting this propagator and putting it on-shell,  the building block has been separated to two trees.  One has  $n_1$ external legs at the lower part (so the whole structure  is the tree  of  $(n_1+2)$-points), while  another one  has $n_2=n_L-n_1$ external legs
at the upper part (so the whole structure  is the tree  of  $(n_2+2)$-points). Using the formula \eref{count-tree-n} we get the number of terms related to this propagator is  $C(n_1+2) C(n-n_1+2)$. Summing over all splitting,  we get the number of terms
for the one-loop building block to be
\bea {\cal B}(n)=\sum_{n_1=0}^n C(n_1+2) C(n-n_1+2).~~~~\label{one-loop-building}\eea

Having the building block, we can count terms for these two types of diagrams in Figure \ref{Fe-1}. For the Type (A), since we ask $n_L, n_R\geq 2$ to avoid  one-loop tadpoles and one-loop massless bubbles,  the total number of terms is given by
\bea {\cal N}_{F;A}(n)=n\sum_{n_L=2}^{n-2} \sum_{n_R=2}^{n-n_L} {\cal B}(n_L) {\cal B}(n_R) C(n-n_L-n_R+2) (n-n_L-n_R+1)~.~~~\label{2loop-FeA} \eea
Let us give a brief explanation of formula \eref{2loop-FeA}. First the factor $n$ comes from the sum over all cyclic orderings. The cyclic sum makes also the two loop momenta $\ell_1, \ell_2$ symmetric in the integrand. Secondly, the sum  is over all possible distributions of $n$ legs into four subsets $n_L, m_u, m_d, n_R$ with  constraints that $n_L\geq 2, n_R\geq 2$ and $m_u, m_d\geq 0$. Thirdly, from the Feynman diagrams, it can be seen that the middle part is just the tree-level amplitude of $(2+m_u+m_d)=(n-n_L-n_R+2)$-points. Furthermore, there are
$(n-n_L-n_R+1)$ ways to distribute to $m_u, m_d$ given $n_L, n_R$, so the contribution from the middle part is given by $C(n-n_L-n_R+2) (n-n_L-n_R+1)$.

For the Type (B), the counting is much simpler. Using the formula for our building block,  we get
\bea {\cal N}_{F;B}(n)& =& n\sum_{n_L=1}^{n-1} {\cal B}(n_L) {\cal B}(n-n_L)-2^3 n C(n)~~~\label{2loop-FeB}  \eea
Now let us explain  formula \eref{2loop-FeB}. First the factor $n$ comes again from the sum over the cyclic orderings. Secondly, to exclude reducible two-loop diagrams, we require $n_L\geq 1, n_R\geq 1$ when we sum
 over all different distributions of $n$ to $n_L$ and $n_R$. Furthermore, There are two special cases
 corresponding to two-loop massless bubbles. One is $n_L=1$ and another one, $n_R=1$. They are multiplying by $C(n)$ because the remaining $(n-1)$-legs must be grouped together to become one.
 The factor $2^3$ is because each massless bubble will produce four trees by different combinations of two cuts, while another two comes from two choices of either $n_L=1$ or $n_R=1$.

 Summing these two parts together, finally we get the number of terms after the partial fraction using expressions from
Feynman diagrams
\bea {\cal N}_F(n)= {\cal N}_{F;B}(n)+{\cal N}_{F;A}(n)~~~\label{Fe-count}\eea

~\\{\bf Comparison:} It can be checked that \eref{Fe-count} is equal to \eref{n-CHY} although they are completely different expressions. The matching serves as a strong consistent check.

\section{Conclusion}

In this paper, we have established the all-loop scattering equations by deforming the loop momenta to higher dimension. Under this new aspect, we have effectively reduced the loop problem to the forward limit of corresponding  tree one. One technical difficulty of this construction is to remove forward singularities of corresponding tree parts. Using the bi-adjoint $\phi^3$ theory, we have demonstrated
how to achieve this goal for two-loop planar integrands. The method is based on a nice understanding of the mapping rule, especially   how to
construct the CHY-integrand which produces tree amplitudes with a fixed pole structure. We have supported our two-loop results of $\phi^3$ theory by matching the number  of terms obtained using two different methods.

Although we have focused on the planar part only in this paper, we think the same idea should work for non-planar part as well as not color ordered loop amplitudes. We believe that our construction should be able to generalize to higher loops, at least for $\phi^3$ theory. Another important thing is to understand how to remove the forward singularities of Yang-Mills and Gravity theories based on our results.

\section*{Acknowledgments}

We would like to thank the early participant of Christian~Baadsgaard, N.~E.~J.~Bjerrum-Bohr, Jacob~L.~Bourjaily, Simon~Caron-Huot and  Poul~H.~Damgaard. We would also like to thank discussions with Song He, Rijun Huang, Qinjun jin, Minxin Luo, Junjie Rao and  suggestions for draft by N.~E.~J.~Bjerrum-Bohr, Poul~H.~Damgaard and Rijun Huang. Last but not least
 we would  like to thank the hospitality of the Niels Bohr International Academy where this project was initiated. This work is supported  by the Qiu-Shi Fund and the Chinese NSF under contracts  No.11135006, No.11125523 and No.11575156.


\end{document}